\documentclass[sigconf]{acmart}

\renewcommand\footnotetextcopyrightpermission[1]{} 
\usepackage{amsmath,amsfonts}
\usepackage{algorithmic}
\usepackage{textcomp}
\usepackage{todonotes}
\usepackage{verbatim}
\usepackage{comment}
\usepackage{nth}
\usepackage{multirow}
\usepackage{listings}
\usepackage[T1]{fontenc}
\usepackage{graphicx}
\usepackage[caption=false,font=footnotesize]{subfig}
\usepackage{fixltx2e}
\usepackage{url}
\usepackage[scaled=.8]{beramono}
\usepackage{xspace}
\usepackage{color}
\usepackage{etoolbox}
\usepackage{booktabs}
\usepackage{caption}
\usepackage{soul}
\newcommand*{\ofp}{OFP\xspace}

\acmDOI{10.1145/3286475.3286477}

\hypersetup{draft}
\begin{document}

\title[On the Applicability of PEBS based Online Memory Access \ \\Tracking for
Heterogeneous Memory Management at Scale] {On the Applicability of PEBS based
Online Memory Access Tracking for Heterogeneous Memory Management at Scale}

\author{Aleix Roca Nonell, Balazs Gerofi\textsuperscript{\ddag},
	Leonardo Bautista-Gomez, \\ Dominique Martinet\textsuperscript{\dag},
	Vicen\c{c} Beltran Querol, Yutaka Ishikawa\textsuperscript{\ddag}}
\affiliation{
  \institution{Barcelona Supercomputing Center, Spain}
  \institution{\textsuperscript{\dag}CEA, France}
  \institution{\textsuperscript{\ddag}RIKEN Center for Computational Science, Japan}
}
\email{{aleix.rocanonell, leonardo.bautista, vbeltran}@bsc.es,
	dominique.martinet@cea.fr,
	{bgerofi, yutaka.ishikawa}@riken.jp}

\begin{abstract}
Operating systems have historically had to manage only a single type of memory
device. The imminent availability of heterogeneous memory devices based on
emerging memory technologies confronts the classic single memory model and
opens a new spectrum of possibilities for memory management. Transparent data
movement between different memory devices based on access patterns of
applications is a desired feature to make optimal use of such devices and to
hide the complexity of memory management to the end user. However, capturing
memory access patterns of an application at runtime comes at a cost, which is
particularly challenging for large-scale parallel applications that may be
sensitive to system noise.

In this work, we focus on the access pattern profiling phase prior to the
actual memory relocation.  We study the feasibility of using Intel's Processor
Event-Based Sampling (PEBS) feature to record memory accesses by sampling at
runtime and study the overhead at scale.  We have implemented a custom PEBS
driver in the IHK/McKernel lightweight multi-kernel operating system, one of
whose advantages is minimal system interference due to the lightweight kernel's
simple design compared to other OS kernels such as Linux. We present the PEBS
overhead of a set of scientific applications and show the access patterns
identified in noise sensitive HPC applications. Our results show that clear
access patterns can be captured with a 10\% overhead in the worst-case and
1\% in the best case when running on up to 128k CPU cores (2,048 Intel Xeon Phi
Knights Landing nodes).  We conclude that online memory access profiling using
PEBS at large-scale is promising for memory management in heterogeneous memory
environments.
\end{abstract}

\ccsdesc[500]{Software and its engineering~Operating systems}

\keywords{high-performance computing, operating systems, heterogeneous memory}

\maketitle

\textcopyright 2018 Association for Computing Machinery.                                                                                         
This is the author's version of the work. It is posted here for your personal use.                                                  
Not for redistribution. The definitive Version of Record was published in https://dl.acm.org/citation.cfm?id=3286477 \linebreak
\hspace*{0.5em}https://doi.org/10.1145/3286475.3286477

\section{Introduction}
\label{sec:intro}

The past decade has brought an explosion of new memory technologies.  Various
high-bandwidth memory types, e.g., 3D stacked DRAM (HBM), GDDR and
multi-channel DRAM (MCDRAM) as well as byte addressable non-volatile storage
class memories (SCM), e.g., phase-change memory (PCM), resistive RAM (ReRAM)
and the recent 3D XPoint, are already in production or expected to become
available in the near future.

Management of such heterogeneous memory types is a major challenge for
application developers, not only in terms of placing data structures into the
most suitable memory but also to adaptively move content as application
characteristics changes in time. Operating system and/or runtime level
solutions that optimize memory allocations and data movement by transparently
mapping application behavior to the underlying hardware are thus highly
desired.

One of the basic requirements of a system level solution is the ability to
track the application's memory access patterns in real-time with low overhead.
However, existing solutions for access pattern tracking are often based on
dynamic instrumentation, which have prohibitive overhead for an online
approach~\cite{luk05pin}. Consequently, system level techniques targeting
heterogeneous memory management typically rely on a two-phase model, where the
application is profiled first, based on which the suggested allocation policy
is then determined~\cite{dullor16data, peng17rthms}.

Intel's Processor Event-Based Sampling (PEBS)~\cite{intel64} is an extension to
hardware performance counters that enables sampling the internal execution
state of the CPU (including the most recent virtual address accessed) and
periodically storing a snapshot of it into main memory. The overhead of PEBS
has been the focus of previous works~\cite{larysch16pebs, akiyama17pebs},
however, not in the context of large-scale high-performance computing (HPC).

The hardware PEBS support provides a number of configuration knobs that control
how often PEBS records are stored and how often the CPU is interrupted for
additional background data processing. Because such disruption typically
degrades performance at scale~\cite{Hoefler:10:Characterizing,
Ferreira:08:Characterizing}, it is important to characterize and understand
this overhead to assess PEBS' applicability for heterogeneous memory management
in large-scale HPC. Indeed, none of the previous studies focusing on PEBS'
overhead we are aware of have addressed large-scale environments.

We have implemented a custom PEBS driver in the IHK/McKernel lightweight
multi-kernel operating system~\cite{bgerofi13pspt, bgerofi16ipdps}. Our
motivation for a lightweight kernel (LWK) is threefold. First, lightweight
kernels are known to be highly noise-free and thus they provide an excellent
environment for characterizing PEBS' overhead.  Second, McKernel has a
relatively simple code-base that enables us to rapidly prototype kernel level
features for heterogeneous memory management and allow direct integration with
our PEBS driver. Our custom driver can be easily configured and enables
fine-grained tuning of parameters that are otherwise not available in the Linux
driver (see Section~\ref{sec:design} for more details).  Finally, the Linux
PEBS driver on the platform we used in this study, i.e., the Oakforest-PACS
machine~\cite{OFP17} based on Intel's Xeon Phi Knight's Landing chip, was not
available.

As the baseline for OS level hierarchy memory management, we aimed at answering
the following questions. What is the overhead of real-time memory accesses
tracking at scale?  What is the trade-off between sampling granularity and the
introduced overhead?  Is it feasible to rely on PEBS for collecting such
information online?

Specifically, in this paper we make the following contributions:

\begin{itemize}
	\item An implementation of a custom PEBS driver in an LWK with the ability
	of fine-tuning its parameters

	\item Systematic evaluation of PEBS' overhead on a number of real HPC
	applications running at large scale

	\item Demonstration of captured memory access patterns as the function of
	different PEBS parameters
\end{itemize}

Previous studies have reported PEBS failing to provide increased accuracy with
reset values (see Section~\ref{sec:pebs}) lower than 1024~\cite{larysch16pebs,
akiyama17pebs} as well as the Linux kernel becoming unstable when performing
PEBS based sampling on high frequency~\cite{olson18nas}.  On up to 128k CPU
cores (2,048 Xeon Phi KNL nodes), we find that our custom driver captures
increasingly accurate access patterns reliably even with very low reset values.
Across all of our workloads, PEBS incurs an overhead of 2.3\% on average with
approximately 10\% and 1\% in the worst and best cases, respectively.

The rest of this paper is organized as follows. We begin by explaining the
background and motivations in Section~\ref{sec:background}.  We describe the
design and implementation of our custom PEBS driver in
Section~\ref{sec:design}.  Our large-scale evaluation is provided in
Section~\ref{sec:eval}.  Section~\ref{sec:related} discusses related work, and
finally, Section~\ref{sec:conclusion} concludes the paper.

\section{Background and Motivation}
\label{sec:background}

This section lays the groundwork for the proposed driver architecture by
providing background information on Intel's Processor Event-Based Sampling
facility~\cite{intel64} and the IHK/McKernel lightweight multi-kernel OS
\cite{bgerofi13pspt, bgerofi16ipdps, bgerofi18ipdps}.

\subsection{Processor Event-Based Sampling}
\label{sec:pebs}

Processor Event-Based Sampling (PEBS) is a feature of some Intel
microarchitectures that builds on top of Intel's Performance Counter Monitor
(PCM).

The PCM facility allows to monitor a number of predefined processor performance
parameters (hereinafter called "events") by counting the number of occurrences
of the specified events\footnote{The exact availability of events depends on
the processor's microarchitecture. However, a small set of "architectural
performance events" remain consistent starting from the Intel Core Solo and
Intel Core Duo generation.} in a set of dedicated hardware registers. When a
PCM counter overflows an interrupt is triggered, which eases the process of
sampling.

PEBS extends the idea of PCM by transparently storing additional processor
information while monitoring a PCM event. However, only a small subset of the
PCM events actually support PEBS. A "PEBS record" is stored by the CPU in a
user-defined memory buffer when a configurable number of PCM events, named
"PEBS reset counter value" or simply "reset", occur. The actual PEBS record
format is microarchitecture dependent, but it generally includes the set of
general-purpose registers.

A "PEBS assist" in Intel nomenclature is the action of storing the PEBS record
into the CPU buffer. When the record written in the last PEBS assist reaches a
configurable threshold inside the CPU PEBS buffer, an interrupt is triggered.
The interrupt handler should process the PEBS data and clear the buffer,
allowing the CPU to continue storing more records. The PCM's overflow interrupt
remains inactive while a PCM event is being used with PEBS.

\subsection{Lightweight Multi-kernels}

\begin{figure}[!htb]
	\centering
	\includegraphics[width=0.4\textwidth]{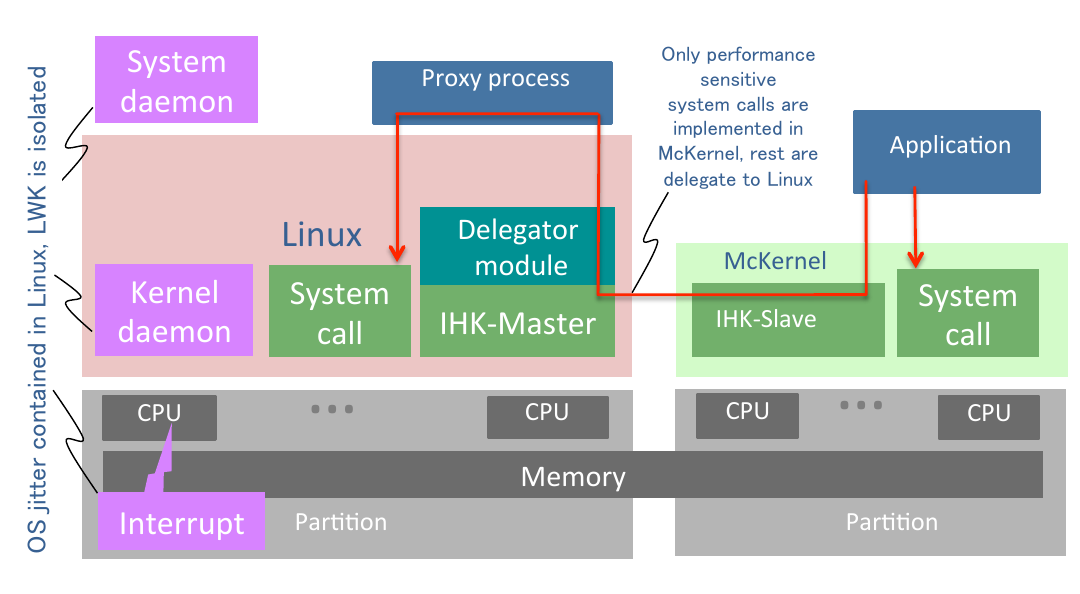}
	\caption{Overview of the IHK/McKernel architecture.}
	\label{fig:ihk+mckernel}
\end{figure}

Lightweight multi-kernels emerged recently as a new operating system
architecture for HPC, where the basic idea is to run
Linux and a LWK side-by-side in compute nodes to attain
the scalability properties of LWKs and full compatibility with Linux at the
same time.  IHK/McKernel is a multi-kernel OS developed at RIKEN, whose
architecture is depicted in Figure \ref{fig:ihk+mckernel}.  A low-level
software infrastructure, called Interface for Heterogeneous Kernels (IHK)
\cite{shimos14ihk}, provides capabilities for partitioning resources in a
many-core environment (e.g., CPU cores and physical memory) and it enables
management of lightweight kernels.  IHK is capable of allocating and releasing
host resources dynamically and no reboot of the host machine is required when
altering its configuration. The latest version of IHK is implemented as a
collection of Linux kernel modules without any modifications to the Linux
kernel itself, which enables relatively straightforward deployment of the
multi-kernel stack on a wide range of Linux distributions.  Besides resource
and LWK management, IHK also facilitates an Inter-kernel Communication (IKC)
layer.

McKernel is a lightweight co-kernel developed on top of IHK.  It is designed
explicitly for HPC workloads, but it retains a Linux
compatible application binary interface (ABI) so that it can execute unmodified
Linux binaries. There is no need for recompiling applications or for any
McKernel specific libraries.  McKernel implements only a small set of
performance sensitive system calls and the rest of the OS services are
delegated to Linux.    Specifically, McKernel provides its own memory
management, it supports processes and multi-threading, it has a simple
round-robin co-operative (tick-less) scheduler, and it implements standard
POSIX signaling. It also implements inter-process memory mappings and it offers
interfaces for accessing hardware performance counters.

For more information on system call offloading, refer to \cite{bgerofi13pspt},
a detailed description of the device driver support is provided in
\cite{bgerofi16ipdps}.  Recently we have demonstrated that lightweight
multi-kernels can indeed outperform Linux on various HPC mini-applications when
evaluated on up to 2,048 Intel Xeon Phi nodes interconnected by Intel's
OmniPath network \cite{bgerofi18ipdps}. As mentioned earlier, with respect to
this study, one of the major advantages of a multi-kernel LWK is the
lightweight kernel's simple codebase that enables us to easily prototype new
kernel level features.

\section{Design and Implementation}
\label{sec:design}

\begin{figure}[!htb]
\centerline{
	\includegraphics[width=0.5\textwidth]{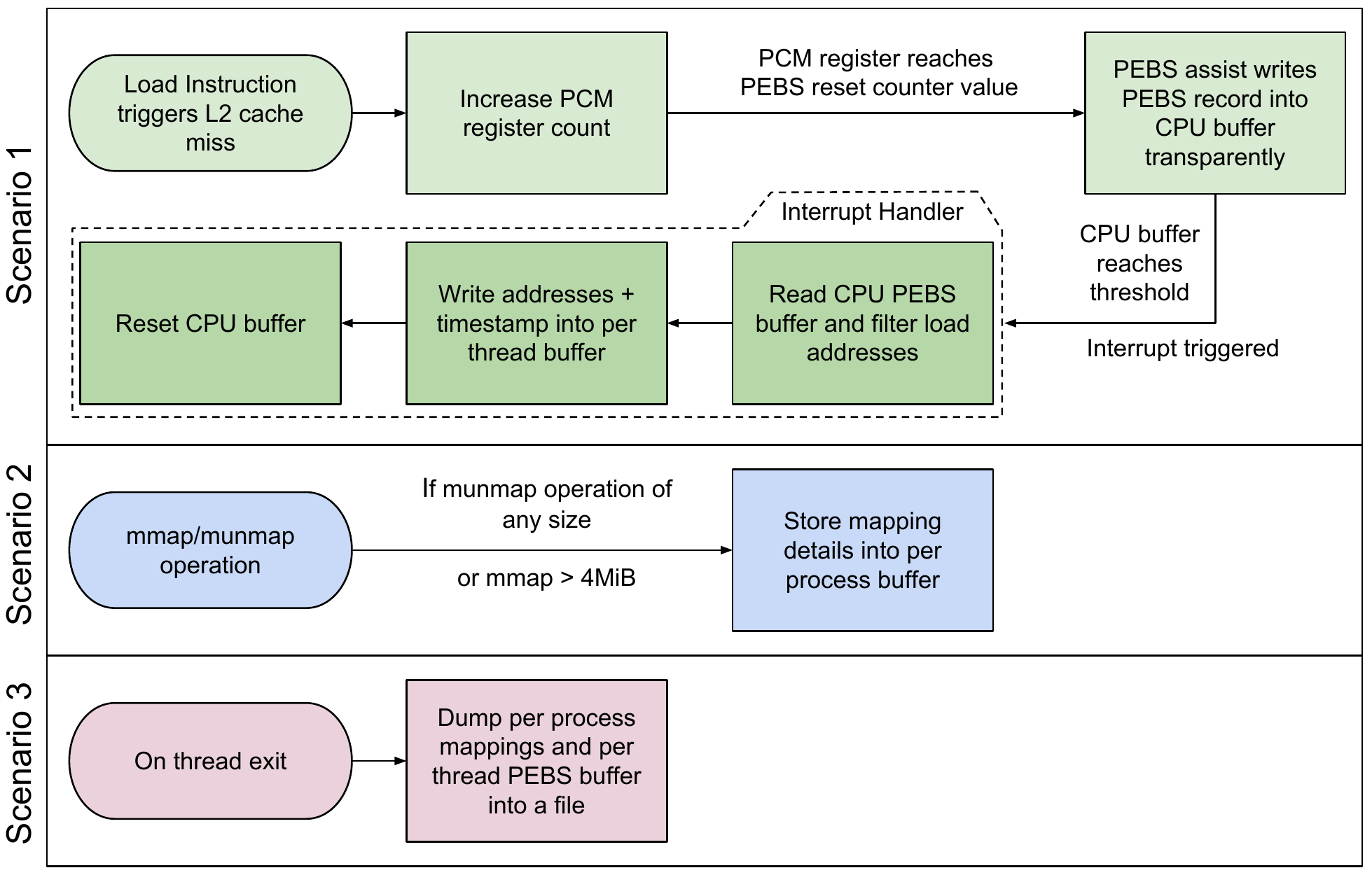}
}
\caption{Memory addresses acquisition processes using Intel's PEBS facility in
IHK/McKernel}
\label{fig:pebs_overview}
\end{figure}

This section describes the design and implementation of the McKernel PEBS
driver. Figure \ref{fig:pebs_overview} shows a summary of the entire PEBS
records lifecycle.

McKernel uses PEBS as a vehicle to keep track of memory addresses issued by
each monitored system thread. Ideally, McKernel would keep track of all load
and store instructions. However, this is not supported by all Intel
microarchitectures. In particular, our test environment powered by the Intel
Knights Landing processor only supports recording the address of load
instructions that triggered some particular event. PEBS records are always
associated with a PCM event. The most general KNL PCM events that support load
address recording are L2\_HIT\_LOADS and L2\_MISS\_LOADS which account for L2
hits and L2 misses, respectively.

Both the count of L2 misses and L2 hits in a page boundary for a given time
frame can be used as a metric that determines how likely is the page to be
accessed in the future. A page with a high count of either L2 misses or L2 hits
reveals that the page is under memory pressure. In the case of misses, we
additionally know that the cache is not able to hold the pages long enough to
be reused. And in the case of hits, we know that either pages are accessed with
high enough frequency to remain in the cache or simply the whole critical
memory range fits into the cache.

In principle, a page with a high L2 miss ratio seems to be a good candidate for
being moved into a faster memory device because missing the L2 in the case of
KNL means that data must be serviced from either main memory or the L2 of
another core. However, the same page might actually have a higher ratio of L2
hits, indicating that another page with a lower hit ratio might benefit still
more from being moved. In consequence, fair judgment should take into
consideration both events. Unfortunately, KNL features a single PCM counter
with PEBS support, which means that sampling both events requires to perform
dynamic switching at runtime.  Nonetheless, the purpose of this work is just a
step behind. Our objective is to focus on the study of a single PEBS enabled
PCM counter at scale. Therefore, for simplicity, we decided to rely on the
L2\_MISS\_LOADS event to record the load addresses.

McKernel initializes the PEBS CPU data structures at boot time on each CPU.
Processes running in McKernel will enable PEBS on all the CPUs where its
threads are running as soon as they start. As long as the threads are being
run, PEBS assists will write PEBS records into the CPU's buffer transparently
regardless of their execution context (user or kernel space).

The PEBS record format for the Knights Landing architecture consists of (among
others) the set of general-purpose registers and the address of the load
instruction causing the record dump (PEBS assist) if applicable. In total, 24
64-bit fields are stored, adding up to a total of 192 bytes for each PEBS
record. There is no timestamp information stored in each PEBS record so it is
not possible to know exactly when the record took place.

When the PEBS remaining capacity reaches the configured threshold, an interrupt
is triggered. The PEBS interrupt handler filters all fields in the PEBS records
but the load address and saves them into a per-thread circular buffer. Then,
the CPU PEBS buffer is reset, allowing the CPU to continue storing records.
Altogether with the load addresses, a timestamp is saved at the time the
interrupt handler is running. This timestamp tags all the PEBS records
processed in this interrupt handler execution for posterior analysis.

When each of the application's threads exit, the entire contents of the
per-thread buffer is dumped into a file. We have developed a small python
visualization tool to read and generate plots based on the information
provided.

The registered load addresses might not belong to application-specific user
buffers but from anywhere in the address space. For offline visualization
purposes we are mostly interested in profiling the application's buffers and
hence, it is convenient to provide some means to filter the undesired
addresses. Load addresses can be sparse, and visualizing the entire address
space of an application to detect patterns might be difficult. It is important
to notice that filtering is not a requirement for online monitoring of high
demanded pages, this is only necessary for visualization.

A simple heuristic to do so is to filter out all addresses of small mappings.
To minimize the impact of filtering, the postprocessing is done offline in our
visualization script. Hence, McKernel only keeps track of all mappings greater
than four megabytes by storing its start addresses, the length and the
timestamp at which the operation completed. All munmap operations are also
registered regardless of its size because they might split a bigger tracked
area. The mappings information are stored into a per-process buffer, shared by
all threads using a lock-free queue. The per-process mappings buffer is also
dumped into the PEBS file at each thread's termination time.

Our PEBS addresses viewer loads the file and reconstructs the processes virtual
memory mappings history based on the mmap and munmap memory ranges and
timestamps. Then, it reads all the registered PEBS load addresses and
classifies them into the right spatial and temporal mapping or discards them if
no suitable mapping is found. Finally, individual plots are shown per mapping.

The PEBS data acquisition rate is controlled by the configurable number of
events that trigger a PEBS assist and the size of the CPU PEBS buffer (which
indirectly controls the threshold that triggers an interrupt). We have added a
simple interface into McKernel to dynamically configure these parameters at
application launch time by resizing the CPU buffer and reconfiguring the PEBS
MSR registers as requested. This differs from the current Linux Kernel driver
in which it is only possible to configure the reset counter value but not the
PEBS buffer size.

It would be ideal to have a big enough CPU buffer to hold all load
addresses the application generates to both reduce the memory movements between
buffers and to suppress the interrupts overhead. However, having a small
interrupt rate also diffuses the time perception of memory accesses because
timestamps are associated with PEBS records in the interrupt handler. Therefore,
this implementation actually requires to set up a proper interrupt rate to
understand the evolution of memory accesses in time. Note that instead of
relying on the interrupt handler to harvest the PEBS CPU buffer, another option
is to dedicate a hardware thread to this task. We plan to implement this option
in the near future.

\section{Evaluation}
\label{sec:eval}

\subsection{Experimental Environment}

All of our experiments were performed on Oakforest-PACS (\ofp), a
Fujitsu built, 25 petaflops supercomputer installed at JCAHPC, managed by The
University of Tsukuba and The University of Tokyo~\cite{OFP17}.  \ofp is
comprised of eight-thousand compute nodes that are interconnected by Intel's
Omni Path network. Each node is equipped with an
Intel\textsuperscript{\textregistered} Xeon Phi\textsuperscript{\texttrademark}
7250 Knights Landing (KNL) processor, which consists of 68 CPU cores,
accommodating 4 hardware threads per core.  The processor provides 16~GB of
integrated, high-bandwidth MCDRAM and it also is accompanied by 96~GB of DDR4
RAM. The KNL processor was configured in Quadrant flat mode;
i.e., MCDRAM and DDR4 RAM are addressable at different physical memory
locations and are presented as separate NUMA nodes to the operating system.

The software environment was as follows. Compute nodes run CentOS 7.4.1708 with
Linux kernel version \texttt{\small 3.10.0-693.11.6}. This CentOS distribution
contains a number of Intel supplied kernel level improvements specifically
targeting the KNL processor that were originally distributed in Intel's XPPSL
package. We used Intel MPI Version 2018 Update 1 Build 20171011 (id: 17941) in
this study.

For all experiments, we dedicated 64 CPU cores to the applications (i.e., to
McKernel) and reserved 4 CPU cores for Linux activities. This is a common
scenario for \ofp users where daemons and other system services run on the
first four cores even in Linux only configuration.

\subsection{Mini-applications}

We used a number of mini-applications from the CORAL benchmark
suite~\cite{CORAL:13:Benchmark} and one developed at the The University of
Tokyo.  Along with a brief description, we also provide information regarding
their runtime configuration.

\begin{itemize}
    \item \textbf{GeoFEM} solves 3D linear elasticity problems in
    simple cube geometries by parallel finite-element method~\cite{nakajima03sc}.
	We used weak-scaling for GeoFEM and ran 16 MPI ranks per node, where each
	rank contained 8 OpenMP threads.

	\item \textbf{HPCG} is the High Performance Conjugate Gradients, which is a
	stand-alone code that measures the performance of basic operations in a
	unified code for sparse matrix-vector multiplication, vector updates, and
	global dot products~\cite{HPCG:17:Benchmark}. We used weak-scaling for HPCG
	and ran 8 MPI ranks per node, where each rank contained 8 OpenMP threads.

	\item \textbf{Lammps} is a classical molecular dynamics code, an acronym
	for Large-scale Atomic/Molecular Massively Parallel
	Simulator~\cite{LAMMPS:17:Benchmark}. We used weak-scaling for Lammps
	and ran 32 MPI ranks per node, where each rank contained four OpenMP threads.

	\item \textbf{miniFE} is a proxy application for unstructured implicit
	finite element codes~\cite{MiniFE:17:Benchmark}. We used strong-scaling for
	miniFE and ran 16 MPI ranks per node, where each rank contained four OpenMP
	threads.

	\item \textbf{Lulesh} is the Livermore Unstructured Lagrangian Explicit
	Shock Hydrodynamics code which was originally defined and as one of five
	challenge problems in the DARPA UHPC program~\cite{LULESH:17:Benchmark}. We
	used weak-scaling for Lulesh and ran 8 MPI ranks per node, where each rank
	contained 16 OpenMP threads.

	\item \textbf{AMG2013} is a parallel algebraic multigrid solver for linear
	systems arising from problems on unstructured grids~\cite{AMG:17:Benchmark}.
	We used weak-scaling for AMG and ran 16 MPI ranks per node, where each
	rank contained 16 OpenMP threads.
\end{itemize}

\subsection{Results}

\begin{figure*}[!htb]
\centerline{
\subfloat[GeoFEM (The University of Tokyo)]{
	\includegraphics[width=0.49\textwidth]{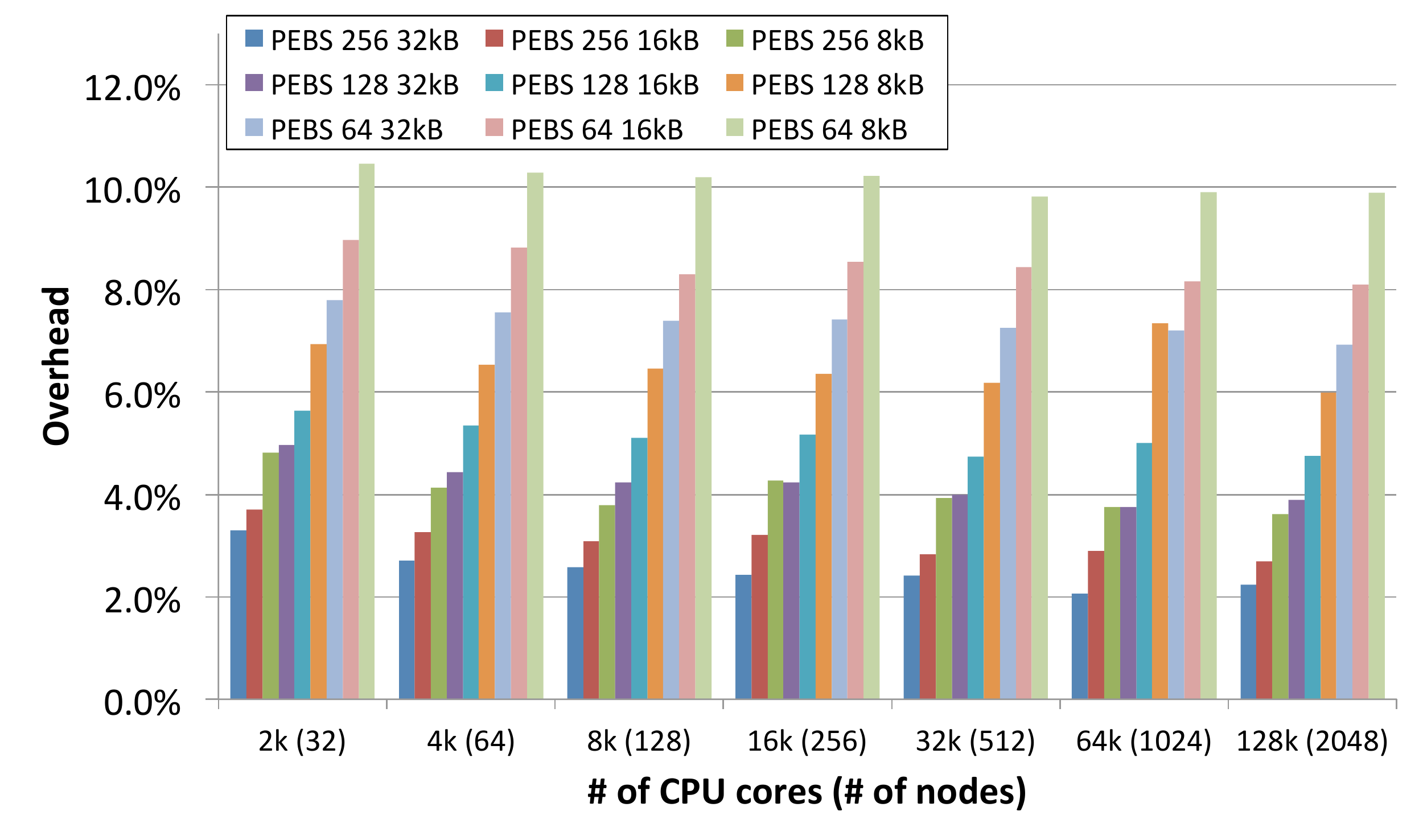}
	\label{fig:geofem}
}
\hspace{0.1cm}
\subfloat[HPCG (CORAL)]{
	\includegraphics[width=0.49\textwidth]{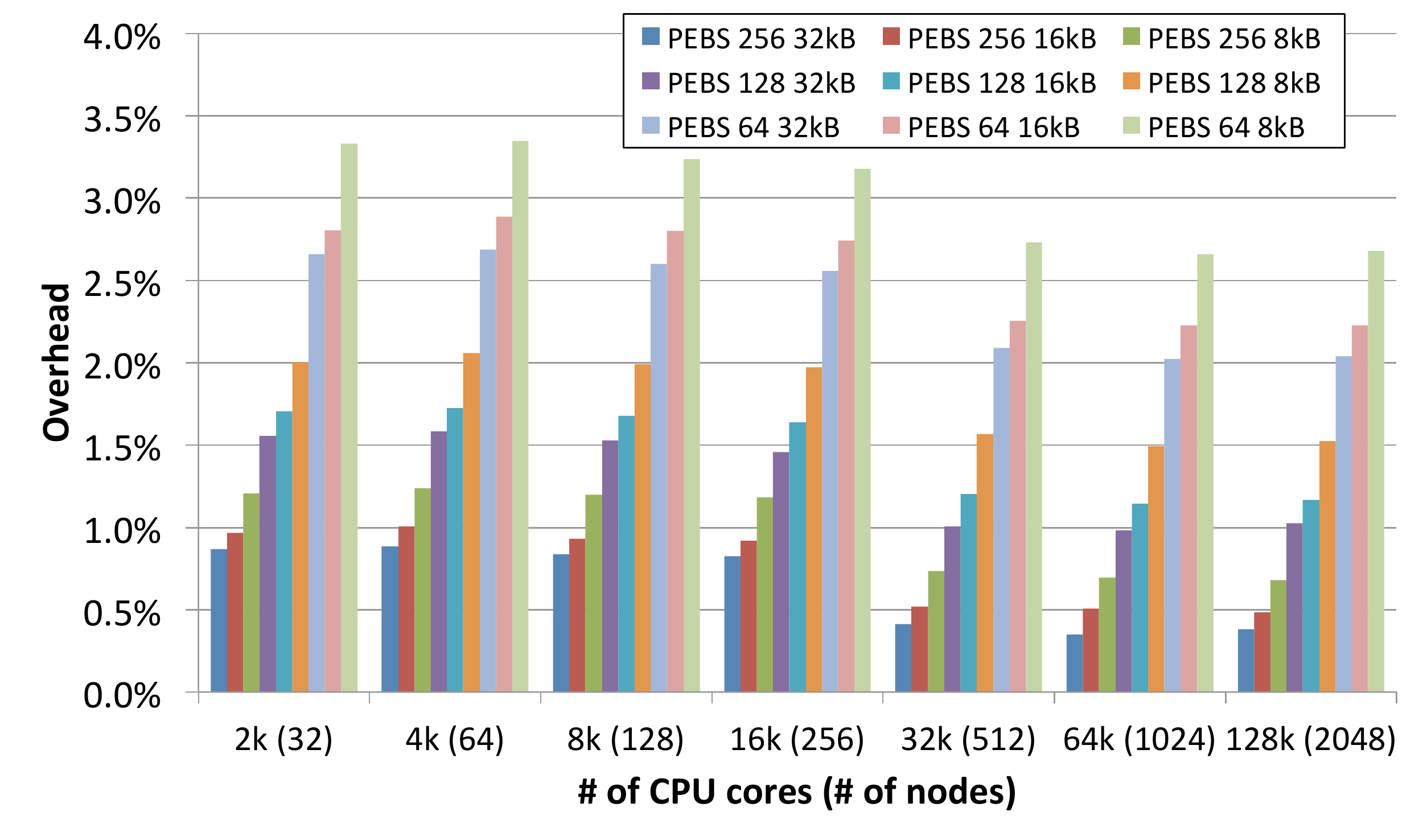}
	\label{fig:hpcg}
}
}
\vspace{0.5cm}
\centerline{
\subfloat[LAMMPS (CORAL)]{
	\includegraphics[width=0.49\textwidth]{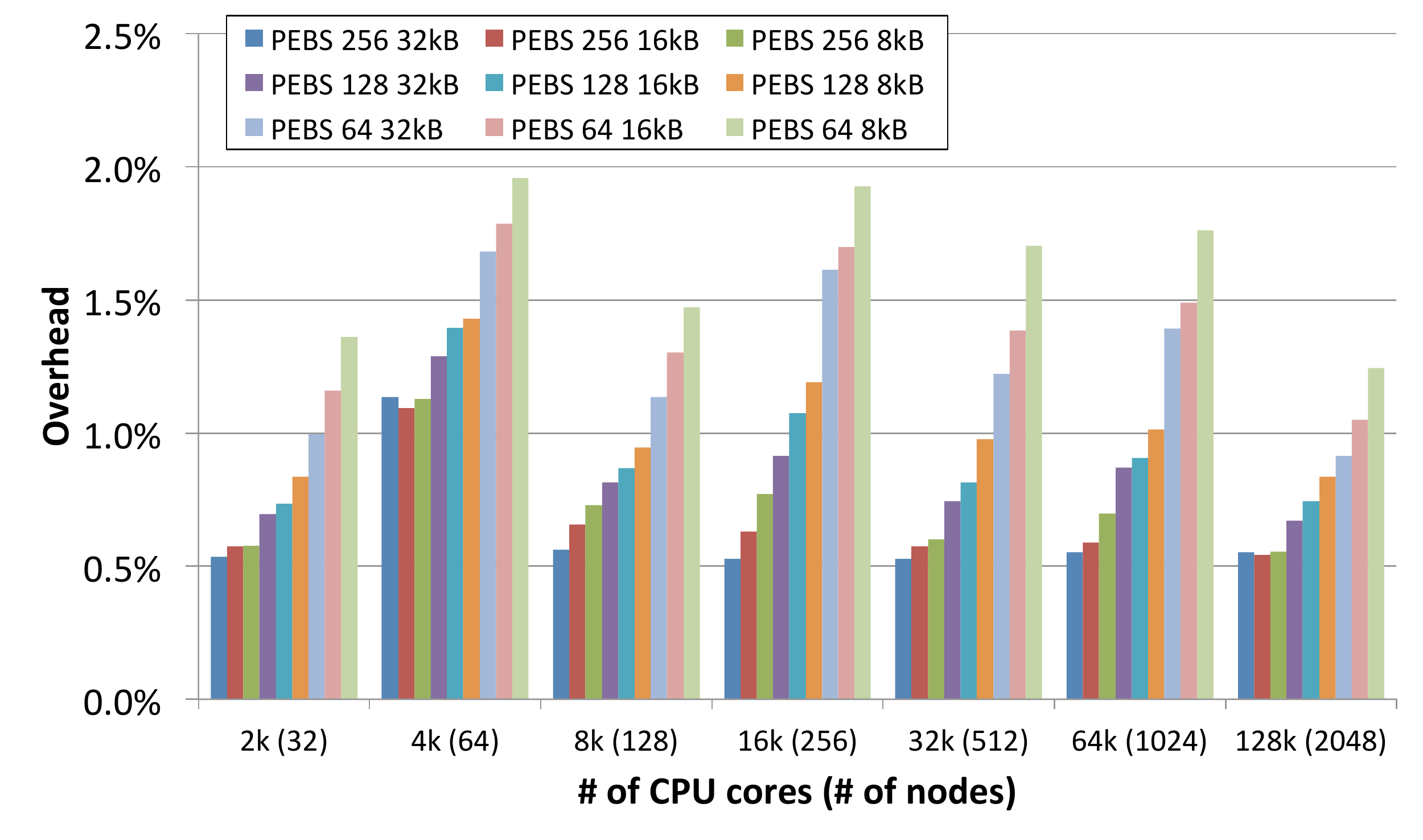}
	\label{fig:lammps}
}
\hspace{0.1cm}
\subfloat[Lulesh (CORAL)]{
	\includegraphics[width=0.49\textwidth]{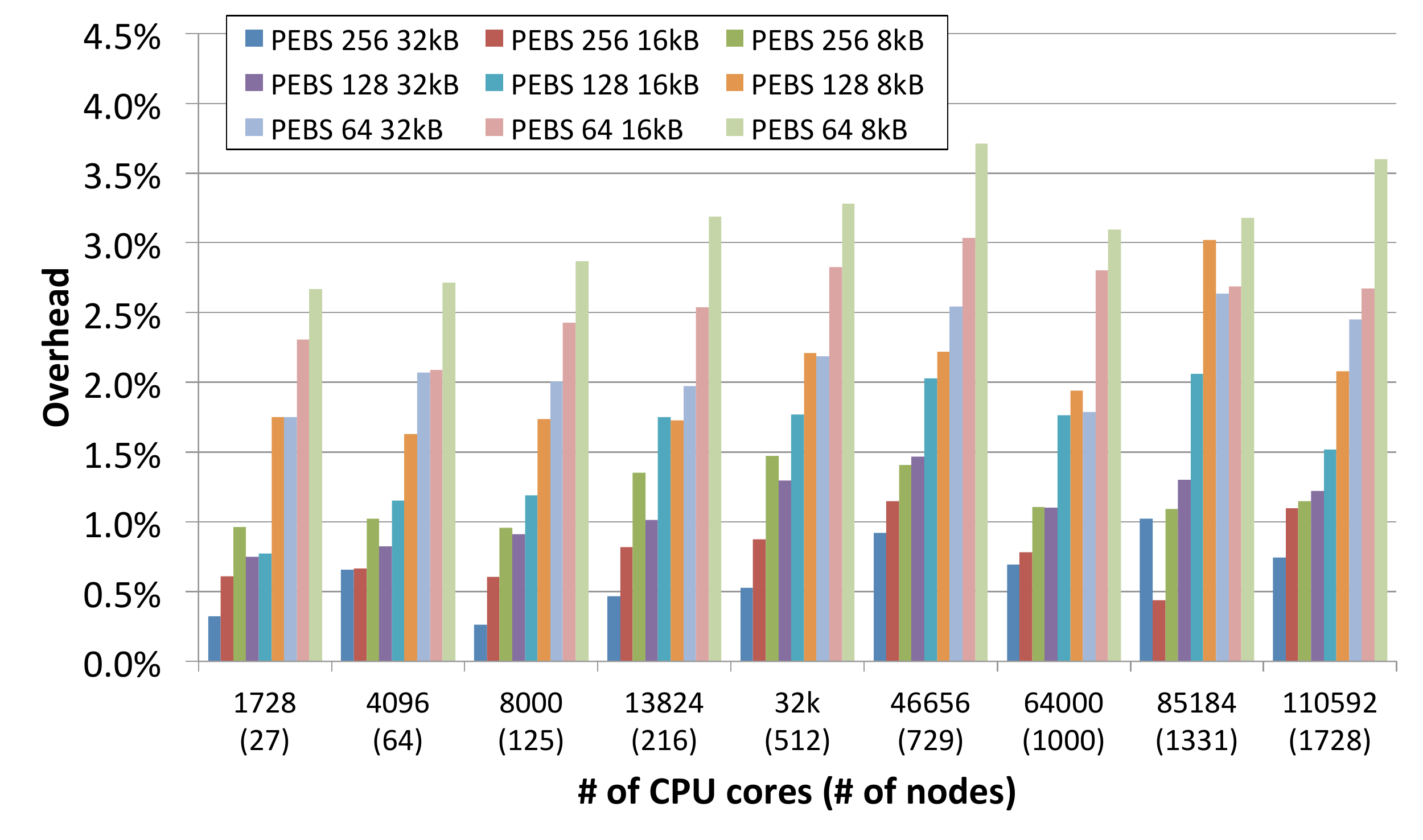}
	\label{fig:lulesh}
}
}
\vspace{0.5cm}
\centerline{
\subfloat[MiniFE (CORAL)]{
	\includegraphics[width=0.49\textwidth]{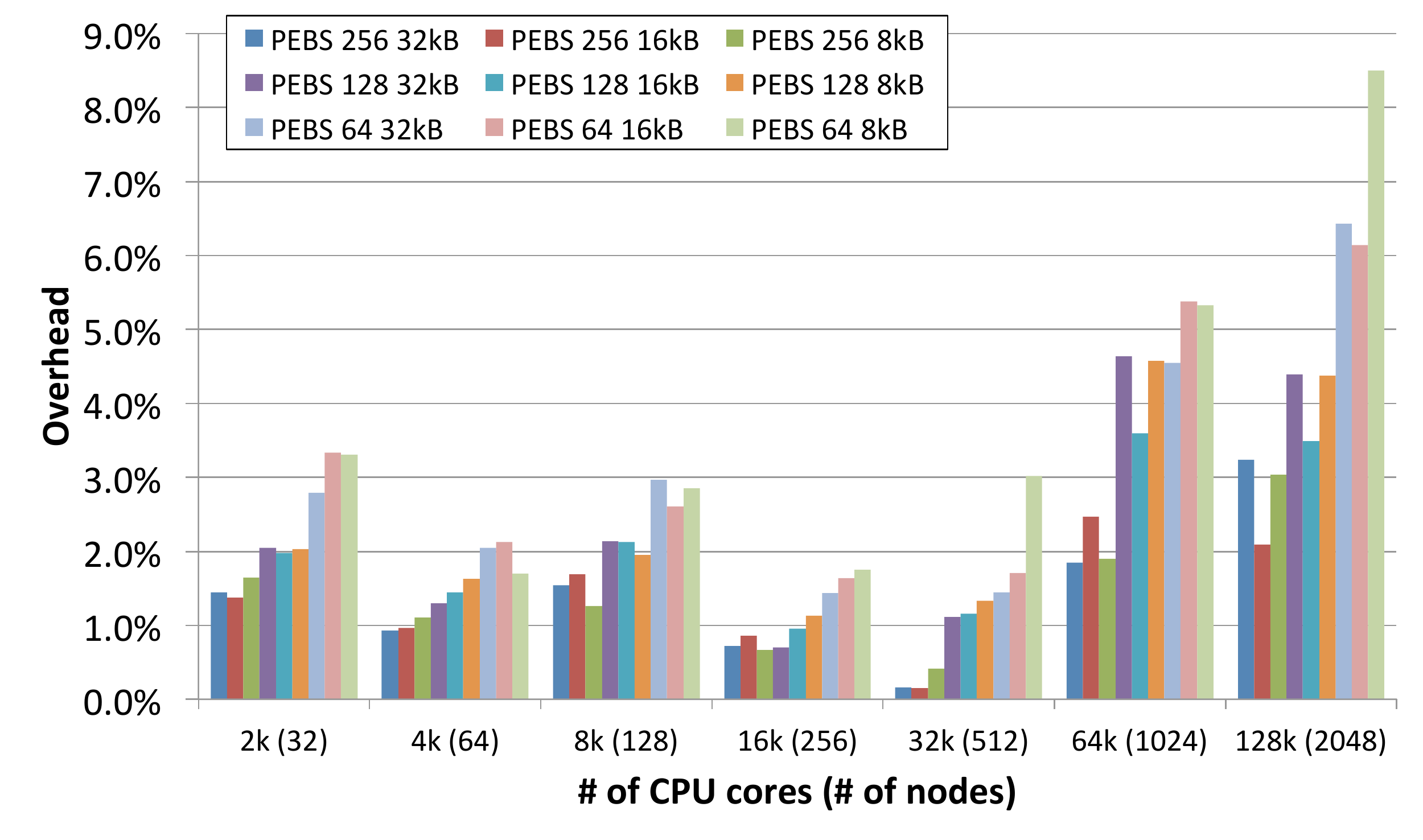}
	\label{fig:minife}
}
\hspace{0.1cm}
\subfloat[AMG2013 (CORAL)]{
	\includegraphics[width=0.49\textwidth]{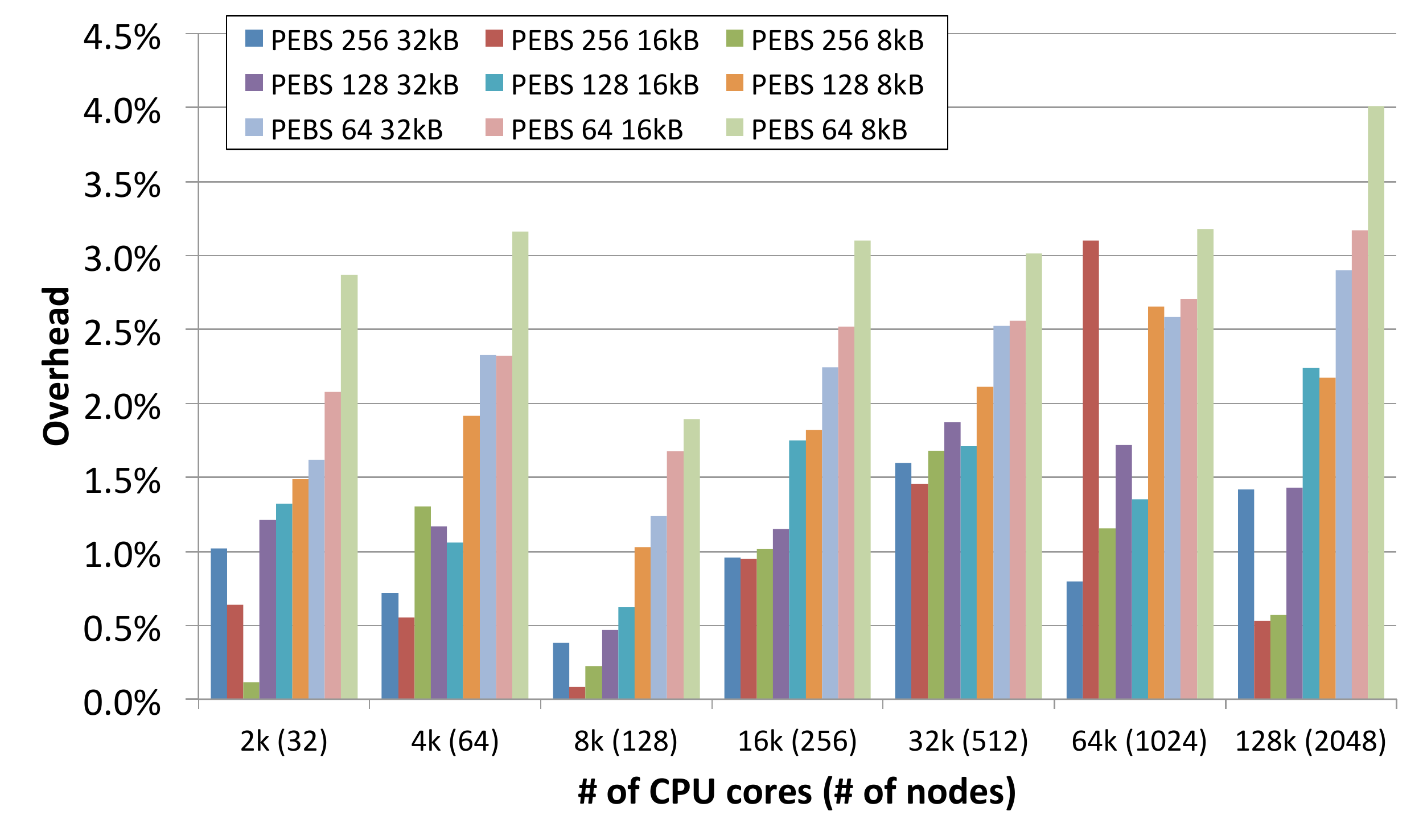}
	\label{fig:amg2013}
}
}
\caption{PEBS overhead for GeoFEM, HPCG, LAMMPS, Lulesh, MiniFE and AMG on up
to 2,048 Xeon Phi KNL nodes}
\label{fig:overhead}
\end{figure*}

\begin{figure*}[!htb]
\centerline{
\subfloat[PEBS reset = 64]{
	\includegraphics[width=0.33\textwidth]{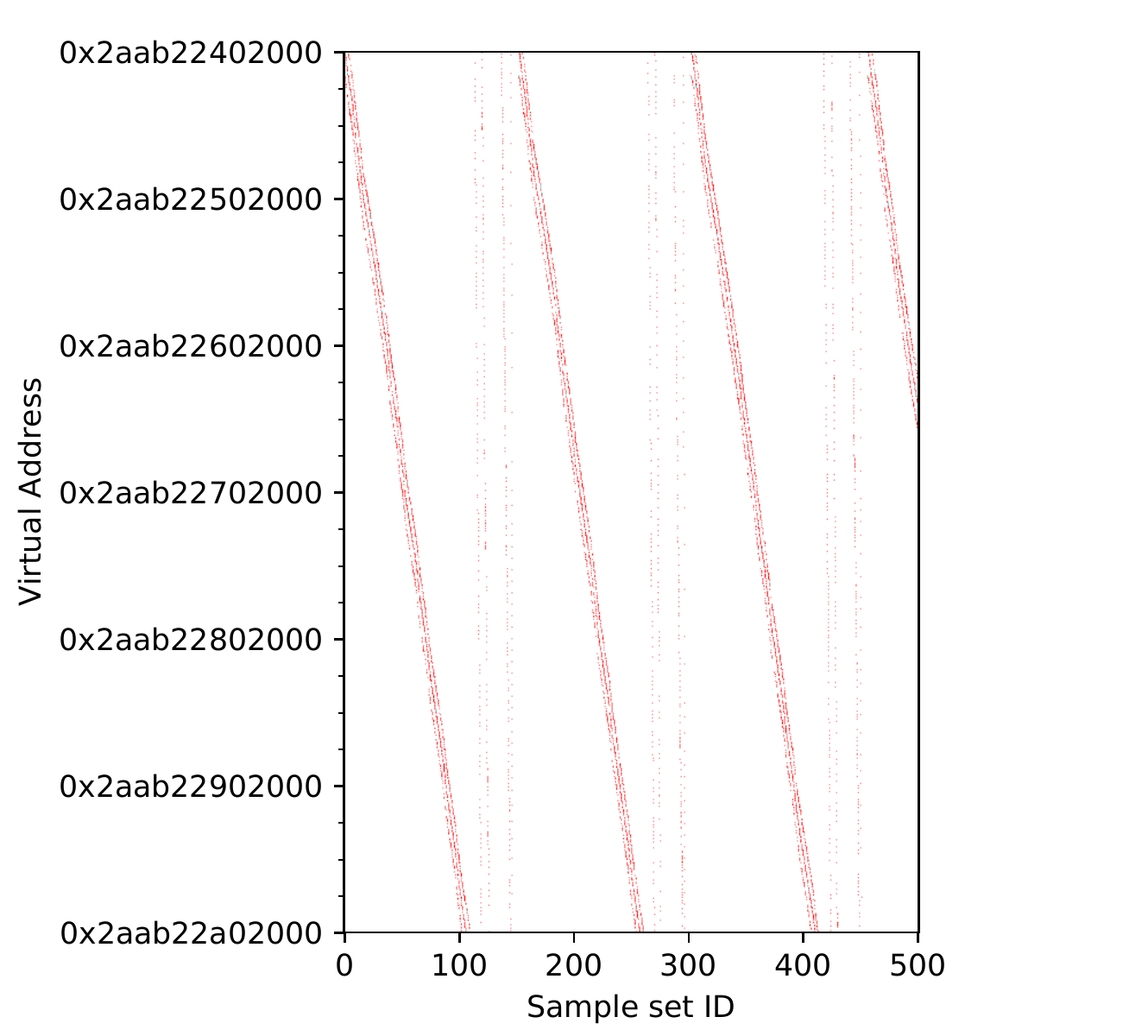}
	\label{}
}
\subfloat[PEBS reset = 128]{
	\includegraphics[width=0.33\textwidth]{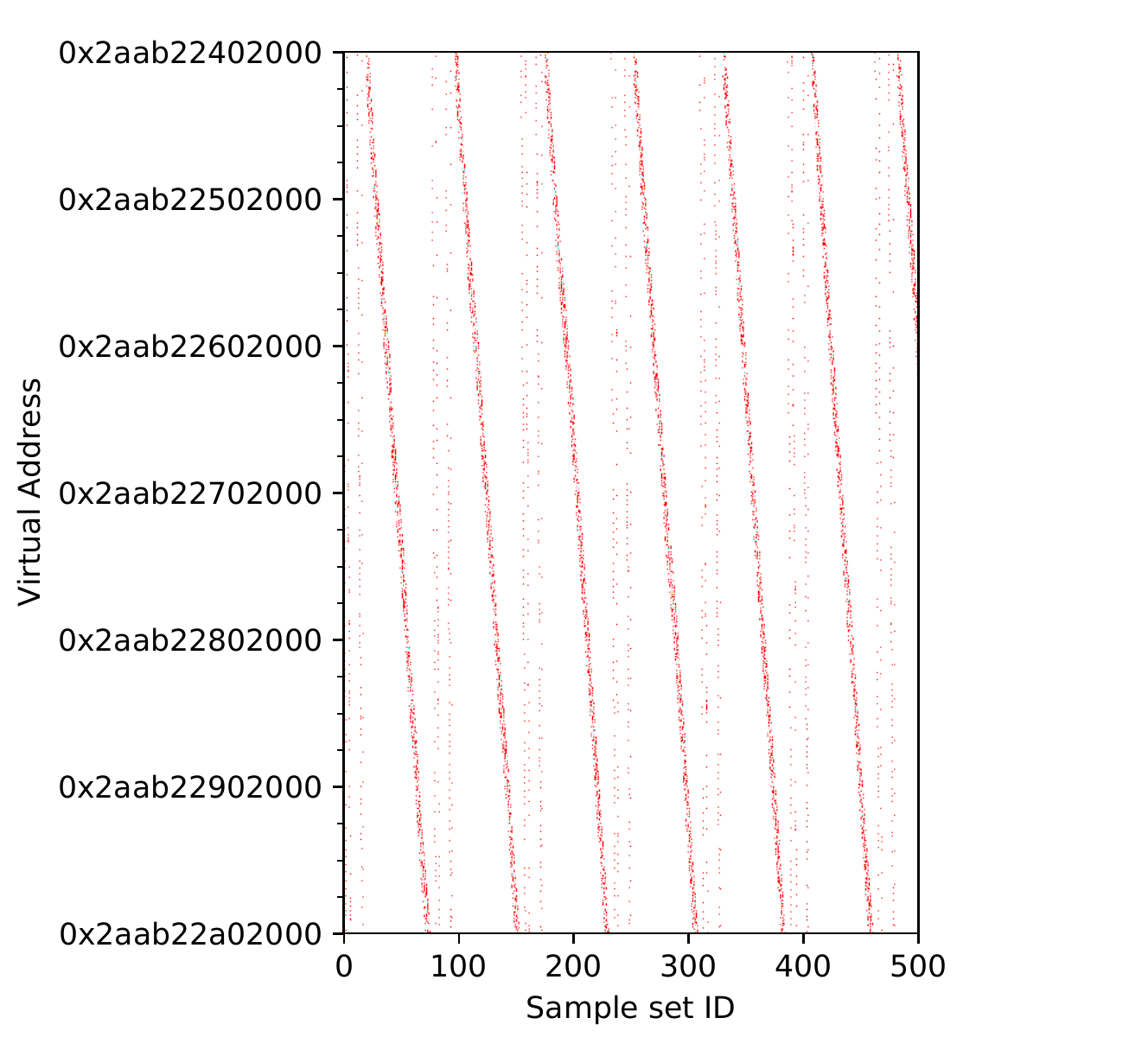}
	\label{}
}
\subfloat[PEBS reset = 256]{
	\includegraphics[width=0.33\textwidth]{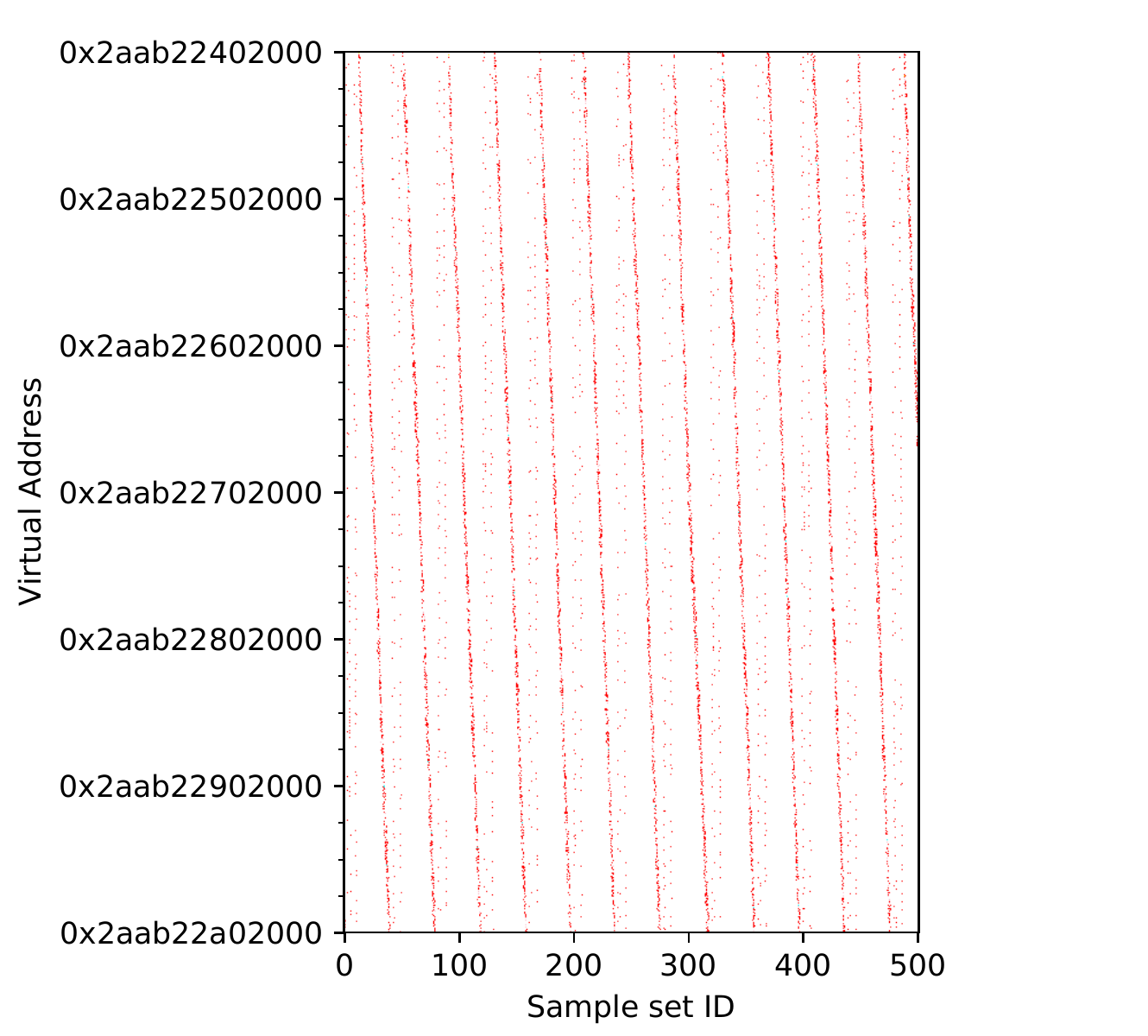}
	\label{}
}
}
\caption{MiniFE access pattern with different PEBS reset values (8kB PEBS
buffer)}
\label{fig:minife-access}
\end{figure*}

\begin{figure*}[!htb]
\centerline{
\subfloat[PEBS reset = 64]{
	\includegraphics[width=0.33\textwidth]{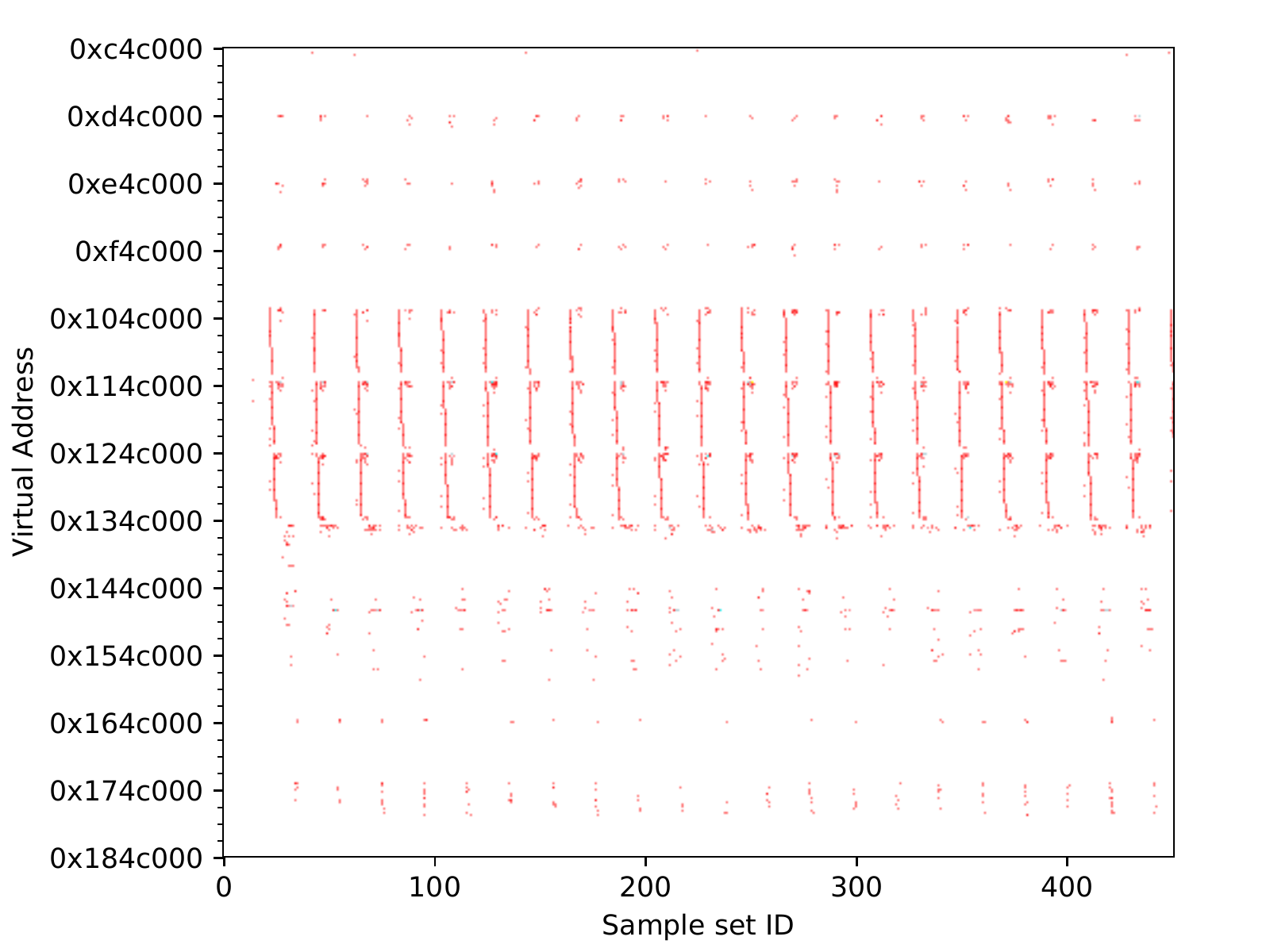}
	\label{}
}
\subfloat[PEBS reset = 128]{
	\includegraphics[width=0.33\textwidth]{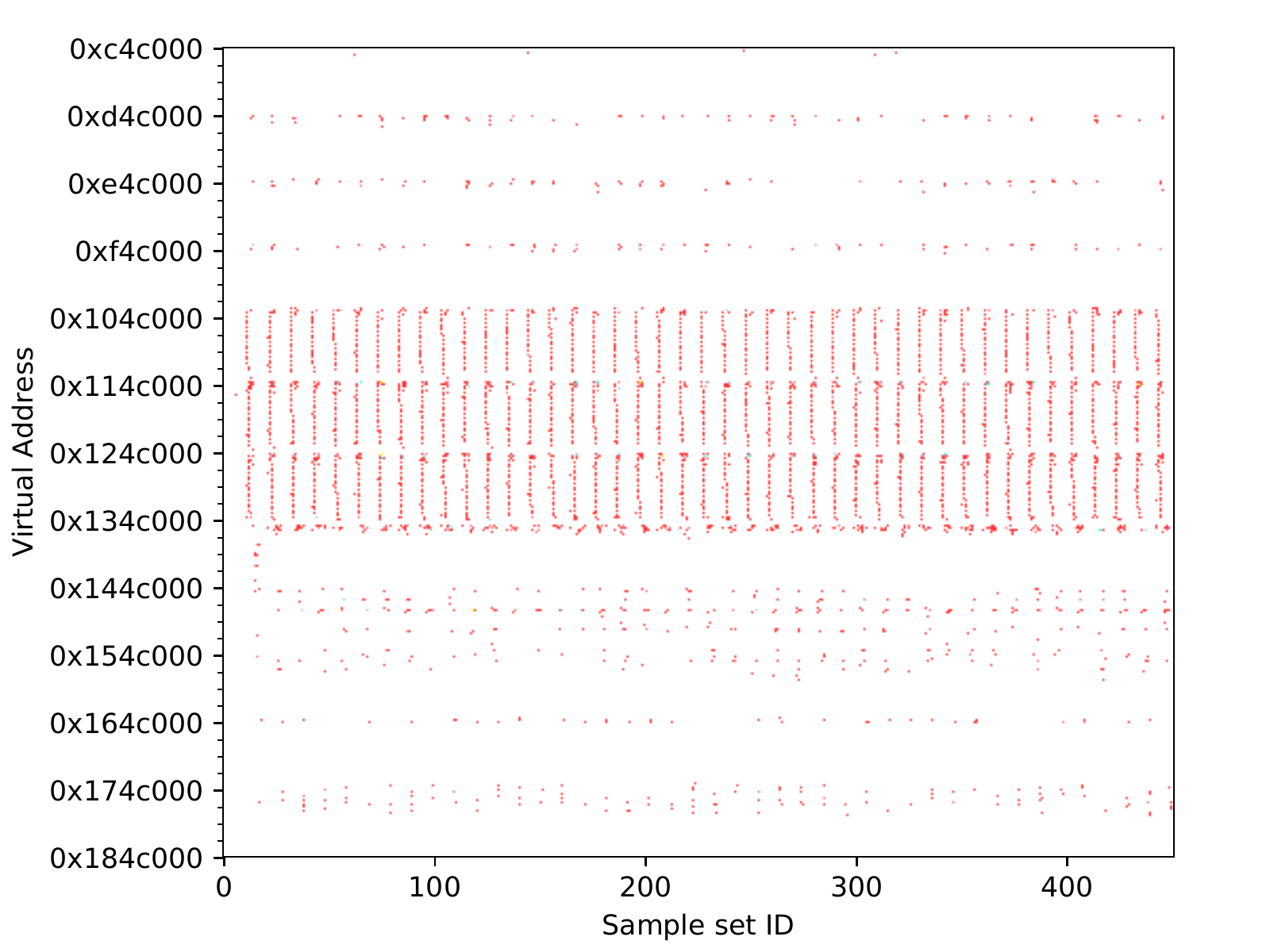}
	\label{}
}
\subfloat[PEBS reset = 256]{
	\includegraphics[width=0.33\textwidth]{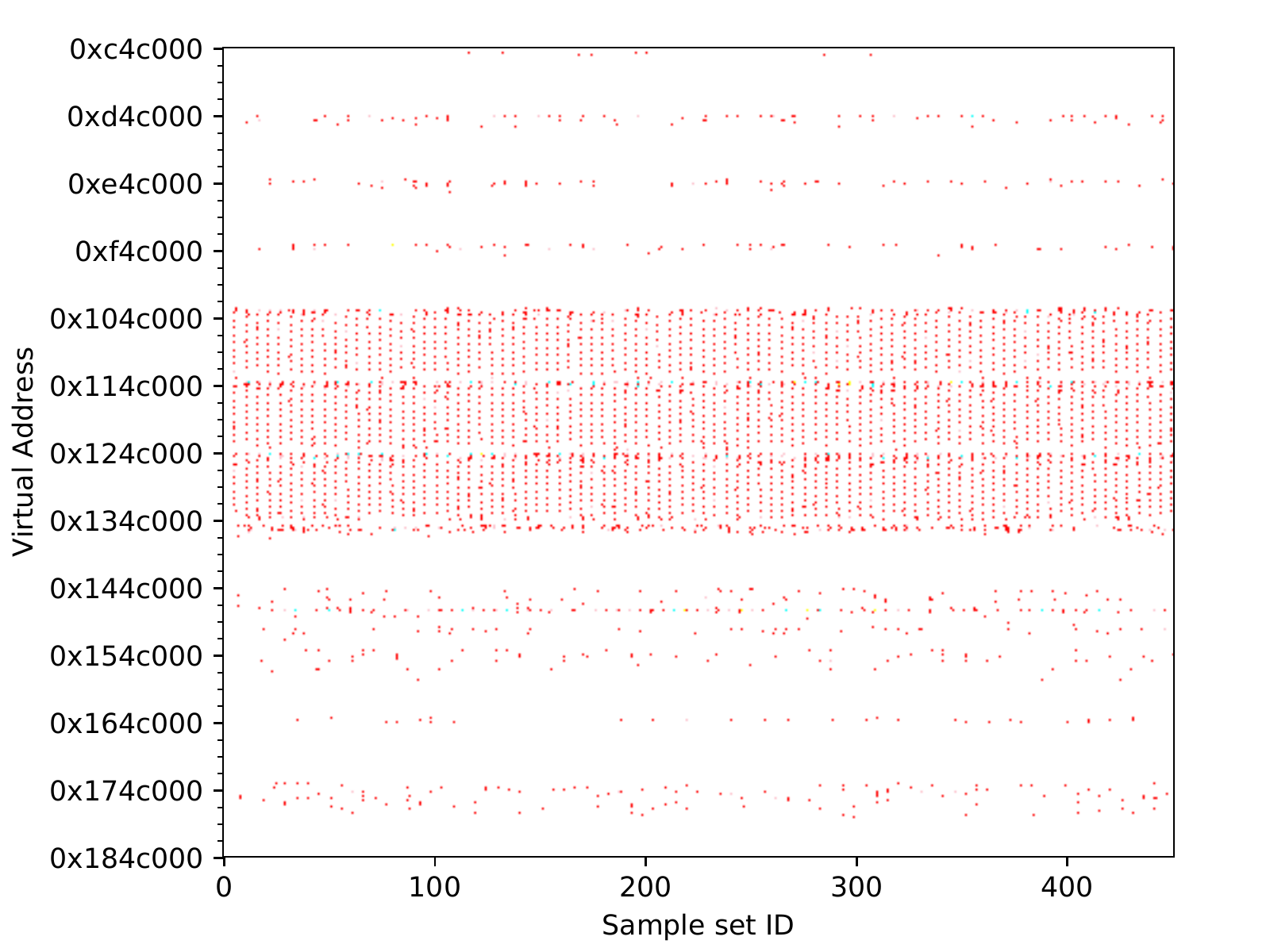}
	\label{}
}
}
\caption{Lulesh access pattern with different PEBS reset values (8kB PEBS
buffer)}
\label{fig:lulesh-access}
\end{figure*}

For each workload described above, we use nine different PEBS
configurations. We scale the PEBS reset value from 256, through 128 to 64 and
used PEBS per-CPU buffer sizes of 8kB, 16kB and 32kB. As mentioned earlier, the
reset value controls the sampling granularity while the PEBS buffer size
impacts the PEBS IRQ frequency.  We emphasize again that contrary to previous
reports on PEBS' inability to provide increased accuracy with reset values
lower than 1024~\cite{larysch16pebs, akiyama17pebs, olson18nas}, we find very
clear indications that obtaining increasingly accurate samples with lower reset
values is possible, for which we provide more information below.

We ran each workload for all configurations scaling from 2,048 to 128k CPU
cores, i.e., from 32 to 2,048 compute nodes, respectively.  We compare
individually the execution time of each benchmark run on McKernel with and
without memory accesses tracking enabled.  We report the average value
of three executions, except for a few long-running experiments, where we took
only two samples (e.g., for GeoFEM). Note that all measurements were taken on
McKernel and no Linux numbers are provided. For a detailed comparison between
Linux and McKernel, refer to~\cite{bgerofi18ipdps}.

Figure~\ref{fig:overhead} summarizes our application level findings. The X-axis
represents node counts while the Y-axis shows relative overhead compared to the
baseline performance. For each bar in the plot, the legend indicates the PEBS
reset value and the PEBS buffer size used in the given experiment.  The general
tendency of overhead for most of the measurements matched our expectations,
i.e., the most influential factor in performance overhead is the PEBS reset
value, whose impact can be relaxed to some extent by adjusting the PEBS buffer
size.

Across all workloads, we observe the largest overhead on GeoFEM (shown in
Figure~\ref{fig:geofem}) when running with the lowest PEBS reset value of 64
and the smallest PEBS buffer of 8kB, where the overhead peaked at 10.2\%.
Nevertheless, even for GeoFEM a less aggressive PEBS configuration, e.g., a
reset value of 256 with 32kB PEBS buffer size induces only up to ~4\% overhead.

To much of our surprise, on most workloads PEBS's periodic interruption of the
application does not imply additional overhead as we scale out with the number
of compute nodes. In fact, on some of the workloads, e.g., HPCG (shown in
Figure~\ref{fig:hpcg}) and Lammps (shown in Figure~\ref{fig:lammps}) we even
observe a slight decrease in overhead for which we have currently no precise
explanation and for which identifying its root cause further investigation is
required. Note that both of these workloads were weak scaled and thus are
presumed to compute on a quasi-constant amount of per-process data irrespective
of scale.

One particular application that did experience growing overhead as the scale
increased is MiniFE, shown in Figure~\ref{fig:minife}. MiniFE was the only
workload we ran in strong-scaled configuration and our previous experience with
MiniFE indicates that it is indeed sensitive to system
noise~\cite{bgerofi18ipdps}. Despite the expectation that due to the decreasing
amount of per-process data at larger node counts the PEBS' overhead would
gradually diminish, the disruption from constant PEBS interrupts appears to
amplify its negative impact.

\begin{figure}[!htb]
\centerline{
	\includegraphics[width=0.4\textwidth]{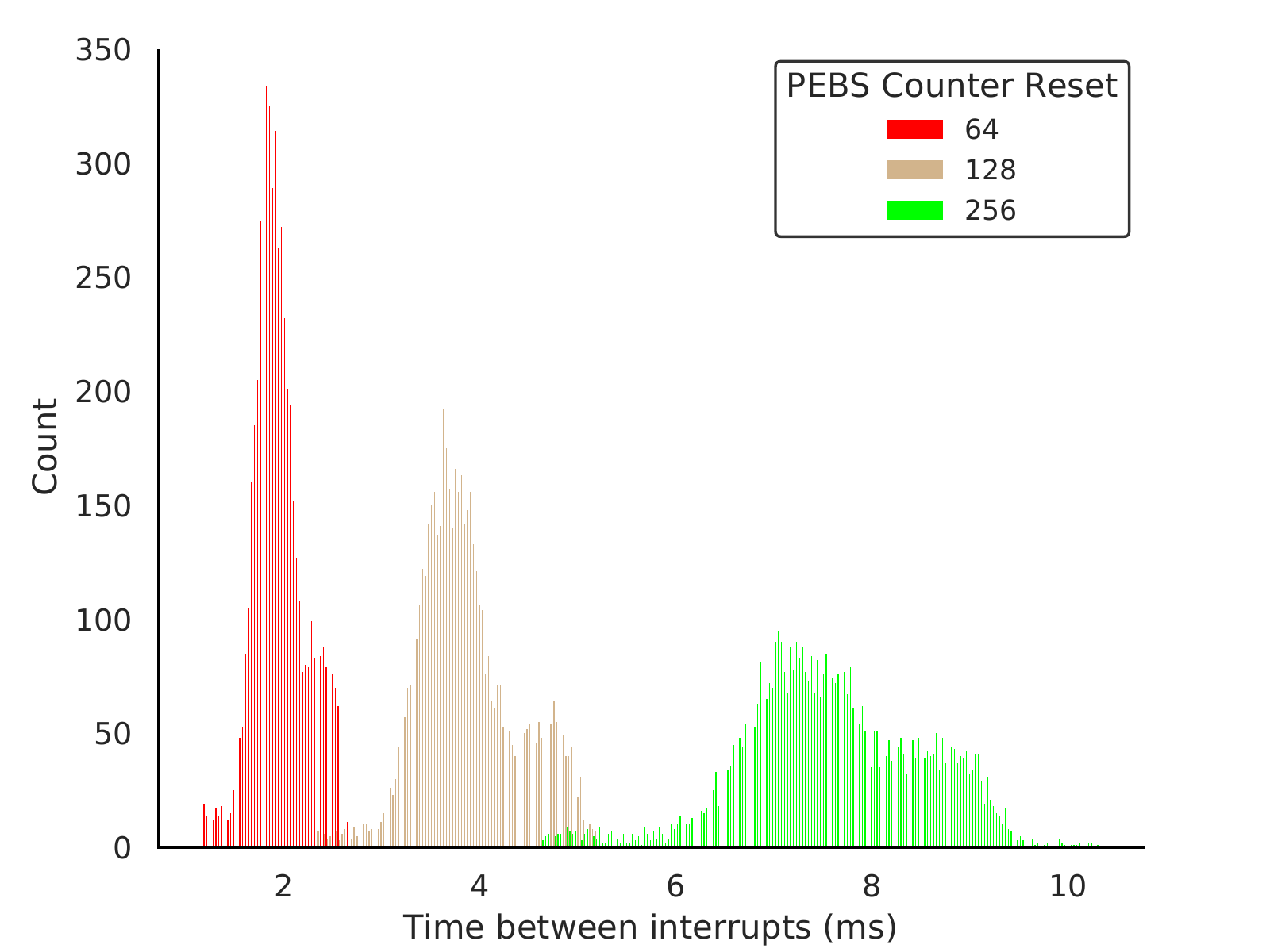}
}
\caption{Distribution of elapsed time between PEBS interrupts for MiniFE with
three different reset values}
\label{fig:int_dist}
\end{figure}

\begin{figure}[!htb]
\centerline{
	\includegraphics[width=0.4\textwidth]{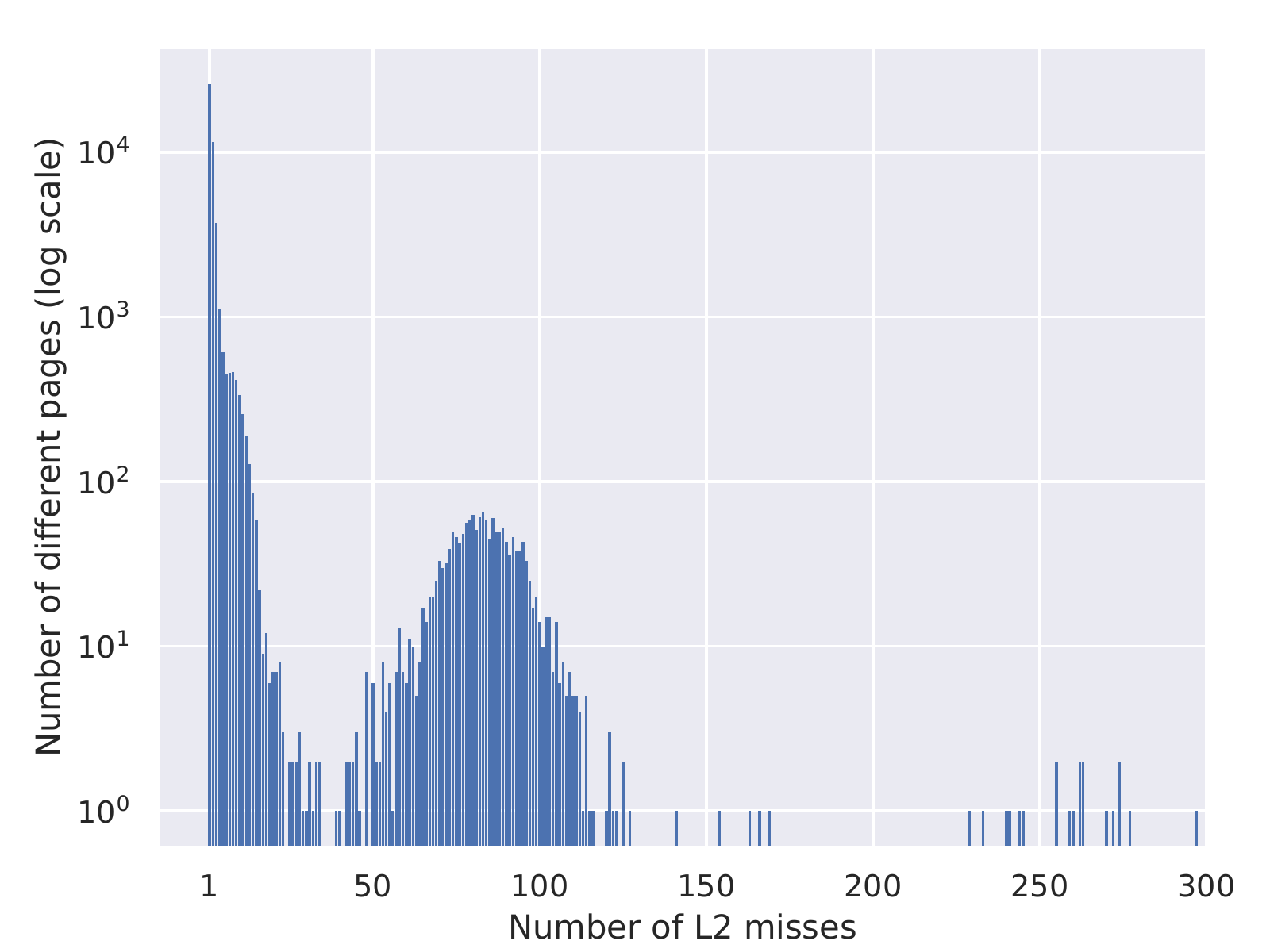}
}
\caption{Access histogram per page for MiniFE execution}
\label{fig:access_hist}
\end{figure}

To demonstrate the impact of PEBS' reset value on the accuracy of memory access
tracking we provide excerpts on memory access patterns using different reset
values. We have been able to observe similar memory access patterns for all
benchmarks tested, but we present the results for MiniFE and Lulesh as an
example. Figure~\ref{fig:minife-access} and Figure \ref{fig:lulesh-access}
show the heatmaps of the access patterns captured on 32 nodes for three reset
values, 64, 128 and 256.  The X-axis represents the sample set ID, i.e.,
periods of time between PEBS interrupts, while the Y-axis indicates the virtual
address of the corresponding memory pages. Although PEBS addresses are captured
at byte granularity, page size is the minimum unit the OS' memory manager works
with.  In fact, for better visibility, we show the heatmap with higher unit
sizes, i.e., in blocks of 4 pages.

One of the key observations here is the increasingly detailed view of the
captured access pattern as we decrease the PEBS reset counter. As seen, halving
the reset value from 128 to 64 gives a 2X higher granularity per sample set,
e.g., the stride access of MiniFE is stretched with respect to the sample IDs.
Note that one iteration of MiniFE's buffer presented in the plot corresponds to
approximately 330ms. To put the accuracy into a more quantitative from the 1536
pages of the buffer shown in the figure, PEBS with 64 reset value reports 1430
pages touched, while using reset values of 128 and 256 report 1157 and 843,
respectively. To the contrary, Lulesh's plots indicate that access patterns
that do not significantly change in time can be captured also with lower
granularity and thus the reset value should be adjusted dynamically based on
the application. Note that the number of computational nodes used affects the
amount of memory each node works with and might alter the visible pattern.
However, as long as the memory share per core does not fit in the L2 the
patterns will generally remain similar.

The implicit effect of altering the PEBS reset counter is the increase or
decrease rate of the PEBS interrupt frequency, assuming a constant workload.
The capacity of controlling the interrupt rate should have a clear impact on
the expected overhead, at least in noise sensitive applications such as minife.
We have presented the relationship between overhead and PEBS reset counter in
Figure \ref{fig:overhead} and we now show the relationship between PEBS reset
counter and interrupt frequency in Figure \ref{fig:int_dist}. The elapsed time
between interrupts is shown for three executions of MiniFE with 64, 128 and 256
values.  As expected, we can see a clear correlation between the average
duration and the reset counter value being the former smaller when the later
decreases. We also note that the duration of the interrupt handler itself took
approximately 20 thousand cycles. It is also interesting to observe the
formation of two close peaks per execution.  This tendency identifies two
different access patterns within the application that lead to a different L2
miss generation scheme.

The presence of particularly hot pages can be easily localized by inspecting
the histogram of aggregated L2 misses shown in Figure \ref{fig:access_hist}.
The plot shows the number of different pages that had N number of L2 misses on
the Y-axis, where N is shown on the X-axis. We can easily see that most of the
pages in MiniFE had a small number of misses at the leftmost side of the
histogram. However, the plot reveals an important group of pages above the 50
L2 misses that could be tagged as movable targets.

In summary, we believe that our large-scale results well demonstrate PEBS'
modest overhead to online memory access tracking and we think that a PEBS based
approach to heterogeneous memory management is worth pursuing.

\section{Related Work}
\label{sec:related}

This section discusses related studies in the domains of heterogeneous
memory management and memory access tracking.

Available tools that help to determine efficient data placement in
heterogeneous memory systems typically require developers to run a profile
phase of their application and modify their code accordingly. Dulloor et
al. proposed techniques to identify data objects to be placed into DRAM in
a hybrid DRAM/NVM configuration~\cite{dullor16data}. Peng et al.
considered the same problem in the context of MCDRAM/DRAM using the Intel
Xeon Phi processor~\cite{peng17rthms}. In order to track memory accesses,
these tools often rely on dynamic instrumentation (such as
PIN~\cite{luk05pin}), which imposes significant performance overhead that
makes it impractical for online access tracking.

Larysch developed a PEBS system to assess memory bandwidth utilization of
applications and reported low overheads, but the authors did not provide a
quantitative characterization of using PEBS for this
purpose~\cite{larysch16pebs}.  Akiyama et al. evaluated PEBS overhead on a set
of enterprise computing workloads with the aim of finding performance anomalies
in high-throughput applications (e.g., Spark, RDBMS)~\cite{akiyama17pebs}. PEBS
has been also utilized to determine data placement in emulated non-volatile
memory based heterogeneous systems~\cite{wu17unimem}.  None of these works,
however, have focused on exclusively studying PEBS overhead on large-scale
configurations. To the contrary, we explicitly target large-scale HPC workloads
to assess the scalability impacts of PEBS based memory access tracking.

Olson et al. reported in a very recent study that decreasing the PEBS reset
value below 128 on Linux caused the system to crash~\cite{olson18nas}.  While
they disclosed results only for a single node setup, we demonstrated that our
custom PEBS driver in McKernel performs reliably and induces low overheads
even when using small PEBS reset values in a large-scale deployment.

\section{Conclusion and Future Work}
\label{sec:conclusion}

This paper has presented the design, implementation and evaluation of a PEBS
driver for the IHK/McKernel which aims to provide the groundwork for an OS
level heterogeneous memory manager. We have shown the captured access patterns
of two scientific applications and demonstrated the evolution of their
resolution as we change the PEBS profiling parameters. We have analyzed the
overhead impact associated with the different recording resolutions in both
timing and interrupt domains at scale up to 128k CPUs (or 2,048 computer nodes)
for six scientific applications. We observed overheads highly dependent on both
the application behavior and the recording parameters which range between 1\%
and 10.2\%.  However, we have been able to substantially reduce the overhead of
our worst-case scenario from 10.2\% to 4\% by adjusting the recording
parameters while still achieving clearly visible access patterns. Our
experience contrast with the current Linux kernel PEBS implementation which is
not capable of achieving very fine-grained sample rates. We conclude that PEBS
efficiency matches the basic requirements to be feasible for heterogeneous
memory management but further work is necessary to quantify the additional
overhead associated with using the recorded data at runtime.

Our immediate future work is to address the challenge of properly using the
recorded addresses at runtime to reorganize memory pages on memory devices
based on access patterns. We will study the benefits of dedicating a hardware
thread to periodically harvest the CPU PEBS buffer instead of relying on
interrupts that constantly pause the execution of the user processes. We also
intend to deeply analyze the difference between the IHK/McKernel PEBS driver
and the Linux kernel driver to better quantify the observed limitations.

\section*{Acknowledgment}

This work has been partially funded by MEXT's program for the Development
and Improvement of Next Generation Ultra High-Speed Computer Systems
under its subsidies for operating the Specific Advanced Large
Research Facilities in Japan.
This project has received funding from the European Union's Horizon 2020
research and innovation programme under the Marie Sklodowska-Curie grant
agreement No 708566 (DURO) and agreement No 754304 (DEEP-EST).

\bibliographystyle{ACM-Reference-Format}
\bibliography{ms}


\begin{thebibliography}{22}


\ifx \showCODEN    \undefined \def \showCODEN     #1{\unskip}     \fi
\ifx \showDOI      \undefined \def \showDOI       #1{#1}\fi
\ifx \showISBNx    \undefined \def \showISBNx     #1{\unskip}     \fi
\ifx \showISBNxiii \undefined \def \showISBNxiii  #1{\unskip}     \fi
\ifx \showISSN     \undefined \def \showISSN      #1{\unskip}     \fi
\ifx \showLCCN     \undefined \def \showLCCN      #1{\unskip}     \fi
\ifx \shownote     \undefined \def \shownote      #1{#1}          \fi
\ifx \showarticletitle \undefined \def \showarticletitle #1{#1}   \fi
\ifx \showURL      \undefined \def \showURL       {\relax}        \fi
\providecommand\bibfield[2]{#2}
\providecommand\bibinfo[2]{#2}
\providecommand\natexlab[1]{#1}
\providecommand\showeprint[2][]{arXiv:#2}

\bibitem[\protect\citeauthoryear{Akiyama and Hirofuchi}{Akiyama and
  Hirofuchi}{2017}]%
        {akiyama17pebs}
\bibfield{author}{\bibinfo{person}{Soramichi Akiyama} {and}
  \bibinfo{person}{Takahiro Hirofuchi}.} \bibinfo{year}{2017}\natexlab{}.
\newblock \showarticletitle{Quantitative Evaluation of {Intel} {PEBS} Overhead
  for Online System-Noise Analysis}. In \bibinfo{booktitle}{\emph{Proceedings
  of the 7th International Workshop on Runtime and Operating Systems for
  Supercomputers ROSS 2017}} \emph{(\bibinfo{series}{ROSS '17})}.
  \bibinfo{publisher}{ACM}, \bibinfo{address}{New York, NY, USA}, Article
  \bibinfo{articleno}{3}, \bibinfo{numpages}{8}~pages.
\newblock
\showISBNx{978-1-4503-5086-0}


\bibitem[\protect\citeauthoryear{{CORAL}}{{CORAL}}{2013}]%
        {CORAL:13:Benchmark}
\bibfield{author}{\bibinfo{person}{{CORAL}}.} \bibinfo{year}{2013}\natexlab{}.
\newblock \bibinfo{title}{Benchmark Codes}.
\newblock \bibinfo{howpublished}{\url{https://asc.llnl.gov/CORAL-benchmarks/}}.
    (\bibinfo{date}{Nov.} \bibinfo{year}{2013}).
\newblock


\bibitem[\protect\citeauthoryear{Corporporation}{Corporporation}{2018}]%
        {intel64}
\bibfield{author}{\bibinfo{person}{Intel Corporporation}.}
  \bibinfo{year}{2018}\natexlab{}.
\newblock \bibinfo{title}{Intel 64 and IA-32 Architectures Software Developer
  Manuals}.
\newblock
  \bibinfo{howpublished}{\url{https://software.intel.com/articles/intel-sdm}}.
   (\bibinfo{year}{2018}).
\newblock


\bibitem[\protect\citeauthoryear{Dongarra, Heroux, and Luszczek}{Dongarra
  et~al\mbox{.}}{2015}]%
        {HPCG:17:Benchmark}
\bibfield{author}{\bibinfo{person}{Jack Dongarra}, \bibinfo{person}{Michael~A.
  Heroux}, {and} \bibinfo{person}{Piotr Luszczek}.}
  \bibinfo{year}{2015}\natexlab{}.
\newblock \bibinfo{booktitle}{\emph{{HPCG} Benchmark: A New Metric for Ranking
  High Performance Computing Systems}}.
\newblock \bibinfo{type}{{T}echnical {R}eport} UT-EECS-15-736.
  \bibinfo{institution}{University of Tennessee, Electrical Engineering and
  Computer Science Department}.
\newblock


\bibitem[\protect\citeauthoryear{Dulloor, Roy, Zhao, Sundaram, Satish,
  Sankaran, Jackson, and Schwan}{Dulloor et~al\mbox{.}}{2016}]%
        {dullor16data}
\bibfield{author}{\bibinfo{person}{Subramanya~R. Dulloor},
  \bibinfo{person}{Amitabha Roy}, \bibinfo{person}{Zheguang Zhao},
  \bibinfo{person}{Narayanan Sundaram}, \bibinfo{person}{Nadathur Satish},
  \bibinfo{person}{Rajesh Sankaran}, \bibinfo{person}{Jeff Jackson}, {and}
  \bibinfo{person}{Karsten Schwan}.} \bibinfo{year}{2016}\natexlab{}.
\newblock \showarticletitle{Data Tiering in Heterogeneous Memory Systems}. In
  \bibinfo{booktitle}{\emph{Proceedings of the Eleventh European Conference on
  Computer Systems}} \emph{(\bibinfo{series}{EuroSys '16})}.
  \bibinfo{publisher}{ACM}, \bibinfo{address}{New York, NY, USA}, Article
  \bibinfo{articleno}{15}, \bibinfo{numpages}{16}~pages.
\newblock
\showISBNx{978-1-4503-4240-7}
\urldef\tempurl%
\url{http://doi.acm.org/10.1145/2901318.2901344}
\showURL{%
\tempurl}


\bibitem[\protect\citeauthoryear{Ferreira, Bridges, and Brightwell}{Ferreira
  et~al\mbox{.}}{2008}]%
        {Ferreira:08:Characterizing}
\bibfield{author}{\bibinfo{person}{Kurt~B. Ferreira}, \bibinfo{person}{Patrick
  Bridges}, {and} \bibinfo{person}{Ron Brightwell}.}
  \bibinfo{year}{2008}\natexlab{}.
\newblock \showarticletitle{Characterizing Application Sensitivity to {OS}
  Interference Using Kernel-level Noise Injection}. In
  \bibinfo{booktitle}{\emph{Proceedings of the 2008 ACM/IEEE Conference on
  Supercomputing}} \emph{(\bibinfo{series}{SC '08})}. \bibinfo{publisher}{IEEE
  Press}, \bibinfo{address}{Piscataway, NJ, USA}, Article
  \bibinfo{articleno}{19}, \bibinfo{numpages}{12}~pages.
\newblock
\showISBNx{978-1-4244-2835-9}


\bibitem[\protect\citeauthoryear{Gerofi, Riesen, Takagi, Boku, Ishikawa, and
  Wisniewski}{Gerofi et~al\mbox{.}}{pear}]%
        {bgerofi18ipdps}
\bibfield{author}{\bibinfo{person}{Balazs Gerofi}, \bibinfo{person}{Rolf
  Riesen}, \bibinfo{person}{Masamichi Takagi}, \bibinfo{person}{Taisuke Boku},
  \bibinfo{person}{Yutaka Ishikawa}, {and} \bibinfo{person}{Robert~W.
  Wisniewski}.} \bibinfo{year}{2018 (to appear)}\natexlab{}.
\newblock \showarticletitle{{Performance and Scalability of Lightweight
  Multi-Kernel based Operating Systems}}. In \bibinfo{booktitle}{\emph{2018
  IEEE International Parallel and Distributed Processing Symposium (IPDPS)}}.
\newblock


\bibitem[\protect\citeauthoryear{Gerofi, Shimada, Hori, and Ishikawa}{Gerofi
  et~al\mbox{.}}{2013}]%
        {bgerofi13pspt}
\bibfield{author}{\bibinfo{person}{Balazs Gerofi}, \bibinfo{person}{Akio
  Shimada}, \bibinfo{person}{Atsushi Hori}, {and} \bibinfo{person}{Yutaka
  Ishikawa}.} \bibinfo{year}{2013}\natexlab{}.
\newblock \showarticletitle{Partially Separated Page Tables for Efficient
  Operating System Assisted Hierarchical Memory Management on Heterogeneous
  Architectures}. In \bibinfo{booktitle}{\emph{13th Intl. Symposium on Cluster,
  Cloud and Grid Computing (CCGrid)}}.
\newblock


\bibitem[\protect\citeauthoryear{Gerofi, Takagi, Hori, Nakamura, Shirasawa, and
  Ishikawa}{Gerofi et~al\mbox{.}}{2016}]%
        {bgerofi16ipdps}
\bibfield{author}{\bibinfo{person}{B. Gerofi}, \bibinfo{person}{M. Takagi},
  \bibinfo{person}{A. Hori}, \bibinfo{person}{G. Nakamura}, \bibinfo{person}{T.
  Shirasawa}, {and} \bibinfo{person}{Y. Ishikawa}.}
  \bibinfo{year}{2016}\natexlab{}.
\newblock \showarticletitle{{On the Scalability, Performance Isolation and
  Device Driver Transparency of the {IHK}/{McKernel} Hybrid Lightweight
  Kernel}}. In \bibinfo{booktitle}{\emph{2016 IEEE International Parallel and
  Distributed Processing Symposium (IPDPS)}}. \bibinfo{pages}{1041--1050}.
\newblock
\showISSN{1530-2075}


\bibitem[\protect\citeauthoryear{Henson and Yang}{Henson and Yang}{2002}]%
        {AMG:17:Benchmark}
\bibfield{author}{\bibinfo{person}{V.~E. Henson} {and} \bibinfo{person}{U.~M.
  Yang}.} \bibinfo{year}{2002}\natexlab{}.
\newblock \showarticletitle{{BoomerAMG}: A Parallel Algebraic Multigrid Solver
  and Preconditioner}.
\newblock \bibinfo{howpublished}{\url{https://codesign.llnl.gov/amg2013.php}}.
\newblock \bibinfo{journal}{\emph{Appl. Num. Math.}}  \bibinfo{volume}{41}
  (\bibinfo{year}{2002}), \bibinfo{pages}{155--177}.
\newblock


\bibitem[\protect\citeauthoryear{Heroux, Doerfler, Crozier, Willenbring,
  Edwards, Williams, Rajan, Keiter, Thornquist, and Numrich}{Heroux
  et~al\mbox{.}}{2009}]%
        {MiniFE:17:Benchmark}
\bibfield{author}{\bibinfo{person}{Michael~A Heroux},
  \bibinfo{person}{Douglas~W Doerfler}, \bibinfo{person}{Paul~S Crozier},
  \bibinfo{person}{James~M Willenbring}, \bibinfo{person}{H~Carter Edwards},
  \bibinfo{person}{Alan Williams}, \bibinfo{person}{Mahesh Rajan},
  \bibinfo{person}{Eric~R Keiter}, \bibinfo{person}{Heidi~K Thornquist}, {and}
  \bibinfo{person}{Robert~W Numrich}.} \bibinfo{year}{2009}\natexlab{}.
\newblock \bibinfo{booktitle}{\emph{{Improving Performance via
  Mini-applications}}}.
\newblock \bibinfo{type}{{T}echnical {R}eport} SAND2009-5574.
  \bibinfo{institution}{Sandia National Laboratories}.
\newblock


\bibitem[\protect\citeauthoryear{Hoefler, Schneider, and Lumsdaine}{Hoefler
  et~al\mbox{.}}{2010}]%
        {Hoefler:10:Characterizing}
\bibfield{author}{\bibinfo{person}{Torsten Hoefler}, \bibinfo{person}{Timo
  Schneider}, {and} \bibinfo{person}{Andrew Lumsdaine}.}
  \bibinfo{year}{2010}\natexlab{}.
\newblock \showarticletitle{Characterizing the Influence of System Noise on
  Large-Scale Applications by Simulation}. In
  \bibinfo{booktitle}{\emph{Proceedings of the 2010 ACM/IEEE International
  Conference for High Performance Computing, Networking, Storage and Analysis}}
  \emph{(\bibinfo{series}{SC '10})}. \bibinfo{publisher}{IEEE Computer
  Society}, \bibinfo{address}{Washington, DC, USA}.
\newblock
\showISBNx{978-1-4244-7559-9}
\urldef\tempurl%
\url{https://doi.org/10.1109/SC.2010.12}
\showDOI{\tempurl}


\bibitem[\protect\citeauthoryear{{Joint Center for Advanced {HPC}
  ({JCAHPC})}}{{Joint Center for Advanced {HPC} ({JCAHPC})}}{2017}]%
        {OFP17}
\bibfield{author}{\bibinfo{person}{{Joint Center for Advanced {HPC}
  ({JCAHPC})}}.} \bibinfo{year}{2017}\natexlab{}.
\newblock \bibinfo{title}{Basic Specification of {Oakforest}-{PACS}}.
\newblock \bibinfo{howpublished}{\url{http://jcahpc.jp/files/OFP-basic.pdf}}.
  (\bibinfo{date}{March} \bibinfo{year}{2017}).
\newblock


\bibitem[\protect\citeauthoryear{Karlin, Keasler, and Neely}{Karlin
  et~al\mbox{.}}{2013}]%
        {LULESH:17:Benchmark}
\bibfield{author}{\bibinfo{person}{Ian Karlin}, \bibinfo{person}{Jeff Keasler},
  {and} \bibinfo{person}{Rob Neely}.} \bibinfo{year}{2013}\natexlab{}.
\newblock \bibinfo{booktitle}{\emph{{LULESH} 2.0 Updates and Changes}}.
\newblock \bibinfo{type}{{T}echnical {R}eport} LLNL-TR-641973.
  \bibinfo{institution}{Lawrence Livermore National Laboratory}.
  \bibinfo{pages}{1--9} pages.
\newblock


\bibitem[\protect\citeauthoryear{Larysch}{Larysch}{2016}]%
        {larysch16pebs}
\bibfield{author}{\bibinfo{person}{Florian Larysch}.}
  \bibinfo{year}{2016}\natexlab{}.
\newblock \emph{\bibinfo{title}{Fine-Grained Estimation of Memory Bandwidth
  Utilization}}.
\newblock Master Thesis. \bibinfo{school}{Operating Systems Group, Karlsruhe
  Institute of Technology (KIT), Germany}.
\newblock


\bibitem[\protect\citeauthoryear{Luk, Cohn, Muth, Patil, Klauser, Lowney,
  Wallace, Reddi, and Hazelwood}{Luk et~al\mbox{.}}{2005}]%
        {luk05pin}
\bibfield{author}{\bibinfo{person}{Chi-Keung Luk}, \bibinfo{person}{Robert
  Cohn}, \bibinfo{person}{Robert Muth}, \bibinfo{person}{Harish Patil},
  \bibinfo{person}{Artur Klauser}, \bibinfo{person}{Geoff Lowney},
  \bibinfo{person}{Steven Wallace}, \bibinfo{person}{Vijay~Janapa Reddi}, {and}
  \bibinfo{person}{Kim Hazelwood}.} \bibinfo{year}{2005}\natexlab{}.
\newblock \showarticletitle{Pin: Building Customized Program Analysis Tools
  with Dynamic Instrumentation}. In \bibinfo{booktitle}{\emph{Proceedings of
  the 2005 ACM SIGPLAN Conference on Programming Language Design and
  Implementation}} \emph{(\bibinfo{series}{PLDI '05})}.
  \bibinfo{publisher}{ACM}, \bibinfo{address}{New York, NY, USA},
  \bibinfo{pages}{190--200}.
\newblock
\showISBNx{1-59593-056-6}


\bibitem[\protect\citeauthoryear{Nakajima}{Nakajima}{2003}]%
        {nakajima03sc}
\bibfield{author}{\bibinfo{person}{Kengo Nakajima}.}
  \bibinfo{year}{2003}\natexlab{}.
\newblock \showarticletitle{Parallel Iterative Solvers of {GeoFEM} with
  Selective Blocking Preconditioning for Nonlinear Contact Problems on the
  {Earth} {Simulator}}. In \bibinfo{booktitle}{\emph{Proceedings of the 2003
  ACM/IEEE Conference on Supercomputing}} \emph{(\bibinfo{series}{SC})}.
  \bibinfo{publisher}{ACM}, \bibinfo{address}{New York, NY, USA}.
\newblock
\showISBNx{1-58113-695-1}
\urldef\tempurl%
\url{https://doi.org/10.1145/1048935.1050164}
\showDOI{\tempurl}


\bibitem[\protect\citeauthoryear{Olson, Zhou, Jantz, Doshi, Lopez, and
  Hernandez}{Olson et~al\mbox{.}}{2018}]%
        {olson18nas}
\bibfield{author}{\bibinfo{person}{Matthew~Benjamin Olson},
  \bibinfo{person}{Tong Zhou}, \bibinfo{person}{Michael~R. Jantz},
  \bibinfo{person}{Kshitij~A. Doshi}, \bibinfo{person}{M.~Graham Lopez}, {and}
  \bibinfo{person}{Oscar Hernandez}.} \bibinfo{year}{2018}\natexlab{}.
\newblock \showarticletitle{MemBrain: Automated Application Guidance for Hybrid
  Memory Systems}. In \bibinfo{booktitle}{\emph{IEEE International Conference
  on Networking, Architecture, and Storage}} \emph{(\bibinfo{series}{NAS'
  18})}. \bibinfo{publisher}{(to appear)}.
\newblock


\bibitem[\protect\citeauthoryear{Peng, Gioiosa, Kestor, Cicotti, Laure, and
  Markidis}{Peng et~al\mbox{.}}{2017}]%
        {peng17rthms}
\bibfield{author}{\bibinfo{person}{Ivy~Bo Peng}, \bibinfo{person}{Roberto
  Gioiosa}, \bibinfo{person}{Gokcen Kestor}, \bibinfo{person}{Pietro Cicotti},
  \bibinfo{person}{Erwin Laure}, {and} \bibinfo{person}{Stefano Markidis}.}
  \bibinfo{year}{2017}\natexlab{}.
\newblock \showarticletitle{RTHMS: A Tool for Data Placement on Hybrid Memory
  System}. In \bibinfo{booktitle}{\emph{Proceedings of the 2017 ACM SIGPLAN
  International Symposium on Memory Management}} \emph{(\bibinfo{series}{ISMM
  2017})}. \bibinfo{publisher}{ACM}, \bibinfo{address}{New York, NY, USA},
  \bibinfo{pages}{82--91}.
\newblock
\showISBNx{978-1-4503-5044-0}


\bibitem[\protect\citeauthoryear{Plimpton}{Plimpton}{1995}]%
        {LAMMPS:17:Benchmark}
\bibfield{author}{\bibinfo{person}{Steve Plimpton}.}
  \bibinfo{year}{1995}\natexlab{}.
\newblock \bibinfo{title}{Fast Parallel Algorithms for Short-range Molecular
  Dynamics}.
\newblock   (\bibinfo{date}{March} \bibinfo{year}{1995}),
  \bibinfo{numpages}{19}~pages.
\newblock
\showISSN{0021-9991}
\urldef\tempurl%
\url{https://doi.org/10.1006/jcph.1995.1039}
\showDOI{\tempurl}


\bibitem[\protect\citeauthoryear{Shimosawa, Gerofi, Takagi, Nakamura,
  Shirasawa, Saeki, Shimizu, Hori, and Ishikawa}{Shimosawa
  et~al\mbox{.}}{2014}]%
        {shimos14ihk}
\bibfield{author}{\bibinfo{person}{Taku Shimosawa}, \bibinfo{person}{Balazs
  Gerofi}, \bibinfo{person}{Masamichi Takagi}, \bibinfo{person}{Gou Nakamura},
  \bibinfo{person}{Tomoki Shirasawa}, \bibinfo{person}{Yuji Saeki},
  \bibinfo{person}{Masaaki Shimizu}, \bibinfo{person}{Atsushi Hori}, {and}
  \bibinfo{person}{Yutaka Ishikawa}.} \bibinfo{year}{2014}\natexlab{}.
\newblock \showarticletitle{Interface for {Heterogeneous} {Kernels}: A
  Framework to Enable Hybrid {OS} Designs targeting High Performance Computing
  on Manycore Architectures}. In \bibinfo{booktitle}{\emph{21th Intl.
  Conference on High Performance Computing}} \emph{(\bibinfo{series}{HiPC})}.
\newblock


\bibitem[\protect\citeauthoryear{Wu, Huang, and Li}{Wu et~al\mbox{.}}{2017}]%
        {wu17unimem}
\bibfield{author}{\bibinfo{person}{Kai Wu}, \bibinfo{person}{Yingchao Huang},
  {and} \bibinfo{person}{Dong Li}.} \bibinfo{year}{2017}\natexlab{}.
\newblock \showarticletitle{Unimem: Runtime Data Managementon Non-volatile
  Memory-based Heterogeneous Main Memory}. In
  \bibinfo{booktitle}{\emph{Proceedings of the International Conference for
  High Performance Computing, Networking, Storage and Analysis}}
  \emph{(\bibinfo{series}{SC '17})}. \bibinfo{publisher}{ACM},
  \bibinfo{address}{New York, NY, USA}, Article \bibinfo{articleno}{58},
  \bibinfo{numpages}{14}~pages.
\newblock
\showISBNx{978-1-4503-5114-0}


\end{thebibliography}

\end{document}